\documentclass[]{aa}
\usepackage[varg]{txfonts}

\usepackage{natbib}
\bibpunct{(}{)}{;}{a}{}{,} 
\usepackage{graphicx}
\usepackage{color}
\usepackage[colorlinks,citecolor=blue]{hyperref}
\usepackage{amstext}
\usepackage{lscape}
\usepackage{multirow}  

\begin{document}

\title{Density diagnostics of ionized outflows in active galactic nuclei}
\subtitle{X-ray and UV absorption lines from metastable levels in Be-like to C-like ions}

\author{Junjie Mao\inst{\ref{inst1},~\ref{inst2}}, 
                    J. S. Kaastra\inst{\ref{inst1},~\ref{inst2}},
            M. Mehdipour\inst{\ref{inst1}},
            A. J. J. Raassen\inst{\ref{inst1},~\ref{inst3}},
            Liyi Gu\inst{\ref{inst1}},
            J. M. Miller\inst{\ref{inst4}}
            }

\institute{SRON Netherlands Institute for Space Research, Sorbonnelaan 2, 
           3584 CA Utrecht, the Netherlands \\
           \email{J.Mao@sron.nl} \label{inst1} 
           \and Leiden Observatory, Leiden University, Niels Bohrweg 2, 
           2300 RA Leiden, the Netherlands \label{inst2}
           \and Astronomical Institute ``Anton Pannekoek", University of Amsterdam, 
           Science Park 904, 1098 XH Amsterdam, the Netherlands \label{inst3}
            \and Department of Astronomy, University of Michigan, 1085 South University Avenue, 
           Ann Arbor, MI 48109-1107, USA \label{inst4}
           }

\date{Received date / Accepted date}

\abstract
{Ionized outflows in active galactic nuclei (AGNs) are thought to influence their nuclear and local galactic environment. However, the distance of the outflows with respect to the central engine is poorly constrained, which limits our understanding of their kinetic power as a cosmic feedback channel. Therefore, the impact of AGN outflows on their host galaxies is uncertain. However, when the density of the outflows is known, their distance can be immediately obtained from their modeled ionization parameters. }
{We perform a theoretical study of density diagnostics of ionized outflows using absorption lines from metastable levels in Be-like to C-like cosmic abundant ions. }
{With the new self-consistent PhotoIONization (PION) model in the SPEX code, we are able to calculate detailed level populations, including the ground and metastable levels. This enables us to determine under what physical conditions the metastable levels are significantly populated. We then identify characteristic lines from these metastable levels in the $1-2000$~\AA\ wavelength range.}
{In the broad density range of $n_{\rm H}\in(10^{6},~10^{20})~{\rm m^{-3}}$, the metastable levels $2s 2p~(^3P_{0-2})$ in Be-like ions can be significantly populated. For B-like ions, merely the first excited level $2s^2 2p~(^2P_{3/2})$ can be used as a density probe. For C-like ions, the first two excited levels $2s^2 2p^2~(^3P_1~{\rm and}~^3P_2)$ are better density probes than the next two excited levels $2s^2 2p^2~(^1S_0~{\rm and}~^1D_{2})$. Different ions in the same isoelectronic sequence cover not only a wide range of ionization parameters, but also a wide range of density values. On the other hand, within the same isonuclear sequence, those less ionized ions probe lower density and smaller ionization parameters. Finally, we reanalyzed the high-resolution grating spectra of \object{NGC\,5548} observed with \textit{Chandra} in January 2002 using a set of PION components to account for the ionized outflow. We derive lower (or upper) limits of plasma density in five out of six PION components based on the presence (or absence) of the metastable absorption lines. Once atomic data from N-like to F-like ions are available, combined with the next generation of spectrometers that cover both X-ray and UV wavelength ranges with higher spectral resolution and larger effective areas, tight constraints on the density and thus the location and kinetic power of AGN outflows can be obtained.}
{}


\titlerunning{Density diagnostics with PION}
\authorrunning{Mao et al.}
\maketitle

\section{Introduction}
\label{sct:intro}
Active galactic nucleus  outflows may play an important role in  cosmic feedback \citep[see, e.g.,][for a review]{fab12}. For an outflow in a thin spherical shell geometry moving constantly with a radial velocity $v$, the mass outflow rate can be estimated via
\begin{equation}
\label{eq:out_rate}
\dot{M}_{\rm out} = 1.43~m_{\rm p}~N_{\rm H}~v~\Omega~ \left(\frac{r^2}{\Delta r}\right),
\end{equation}
where $m_{\rm p}$ is the proton mass, $N_{\rm H}$ the hydrogen column density along the line of sight, $r$ the distance between the outflow and the central engine, $\Delta r$ the radial size of the outflow, and $\Omega$ the solid angle subtended by the outflow. The kinetic power carried by the outflow is $L_{\rm KE} = \frac{1}{2} \dot{M}_{\rm out} v^2$. Thus, distant and/or high-velocity outflow leads to higher mass outflow rate ($\dot{M}_{\rm out}$) and kinetic power ($L_{\rm KE}$). 

While the line-of-sight hydrogen column density ($N_{\rm H}$) and velocity ($v$) of the outflow can be well constrained via spectral analysis, its solid angle ($\Omega$) and location ($r$) are the main source of uncertainties in estimating the kinetic power that impacts the host galaxy. The solid angle highly depends on the exact geometry, which is not investigated here. On the other hand, the location can be immediately obtained via the definition of the ionization parameter\footnote{Throughout this paper, $\xi$ is in units of $10^{-9}~{\rm W~m}$ (i.e., ${\rm erg~s^{-1}~cm}$).} \citep{tar69, kro81} if the density of the outflow is known,
\begin{equation}
\label{eq:xi_def}
\xi = \frac{L}{n_{\rm H}~r^2}
,\end{equation} 
where $L$ is the 1--1000 Ryd (or 13.6 eV--13.6 keV) band luminosity of the ionizing source, $n_{\rm H}$ the hydrogen number density of the ionized plasma, and $r$ the distance of the plasma with respect to the ionizing source. 

However, it is not trivial to determine the density of a photoionized plasma. Three different approaches have been used to measure the density of AGN outflows. The first approach is a timing analysis where  the response of the ionized outflow to changes in the ionizing continuum is monitored. A high-density plasma recombines more rapidly, thus yields a shorter recombination timescale. This approach has been used to constrain density in Mrk\,509 \citep{kaa12}, NGC\,5548 \citep{ebr16}, and NGC\,4051 \citep{sil16}, etc. The timing analysis is in general challenging because the observations of a given object are often sparse, washing out the  possible effects of variability, and lack the signal-to-noise ratio required to significantly measure the expected changes \citep{ebr16}. 

The second approach is a spectral analysis of density sensitive emission lines. It is well known that the ratio of intercombination to forbidden emission lines in the He-like triplets \citep[e.g.,][]{por10}  varies for plasmas with different density values. This density probe, observed in the X-ray wavelength range, has been applied to a few  AGNs, e.g., NGC\,4051 \citep{col01}, NGC\,4593 \citep{mck03}, and NGC\,4151 \citep{sch04}, where the upper limits of the plasma density are derived. Meanwhile, as shown in \citet{meh15a}, line absorption of He-like ion triplet lines by Li-like ions make density diagnostics complicated. In addition, for solar corona studies, metastable emission lines observed in the EUV wavelength range from Be-like \citep{lan14}, B-like \citep{kee98, cia01, gal99, war09}, C-like \citep{kee93, lan98}, and N-like \citep{kee04} ions are widely used to determine the density. 

The third approach is a spectral analysis where density sensitive metastable absorption lines are identified. This method has been successfully used for absorption lines observed in the UV band \citep[e.g.,][]{ara15}. In the X-ray band, \citet{kaa04} obtain an upper limit of the outflow density in Mrk\,279, with the density sensitive metastable absorption transitions $2s^2$--$1s 2s^2 2p$ in \ion{O}{V} (Be-like) $\sim$22.4~\AA. Later, \citet{kin12} reported an upper limit of the outflow density in NGC\,4051 by using the ground (11.77~\AA) and metastable (11.92~\AA) transitions in \ion{Fe}{XXII} (B-like). The same transitions in \ion{Fe}{XXII} were in fact previously used by \citet{mil08} in the stellar mass black hole GRO\,J1655--40, yielding a tight constraint on the density of a disk wind. 

In this work, we carry out a systematic study of density diagnostics with ions in different isoelectronic sequences in photoionized equilibrium. A detailed calculation is made with our new photoionized plasma model PION\footnote{An introduction of PION and a comparison to other photoionization codes can be found in \citet{meh16}} in the SPEX code \citep{kaa96}.  Given cosmic abundances \citep[the proto-solar abundance table in][]{lod09}, we consider elements including C, N, O, Ne, Mg, Si, S, Ar, Ca, and Fe. Throughout this paper we consider plasma densities in the range between $10^6~{\rm m^{-3}}$ (or $1~{\rm cm^{-3}}$) and $10^{20}~{\rm m^{-3}}$ (or $10^{14}~{\rm cm^{-3}}$). At $n_{\rm H}=10^{20}~{\rm m^{-3}}$, metastable levels in Be-like to F-like ions can be significantly populated compared to the population of the ground level, while for Li-like, Ne-like, and Na-like ions, no metastable levels are significantly populated. Therefore, we focus on Be-like to C-like ions in Section~\ref{sct:04ies} to \ref{sct:06ies}. Due to the lack of atomic data, for N-like to F-like ions, we merely discuss \ion{Fe}{XX}, \ion{Fe}{XIX,} and \ion{Fe}{XVIII} in Appendix~\ref{sct:07to09ies}.

\section{Methods}
\label{sct:mo}
Unless specified otherwise, we use the spectral energy distribution (SED) of NGC\,5548 in our photoionization modeling, i.e., the AGN1 SED shown in Fig.~1 of  \citet{meh16}. The photoionized outflow is assumed to be optically thin with a slab geometry. The line-of-sight hydrogen column density ($N_{\rm H}$) is $10^{24}~{\rm m^{-2}}$. The covering factor is unity. No velocity shift with respect to the central engine is assumed here, and the turbulent velocity is set to $100~{\rm km~s^{-1}}$ in the calculation. 

\begin{figure}
\centering
\includegraphics[width=\hsize]{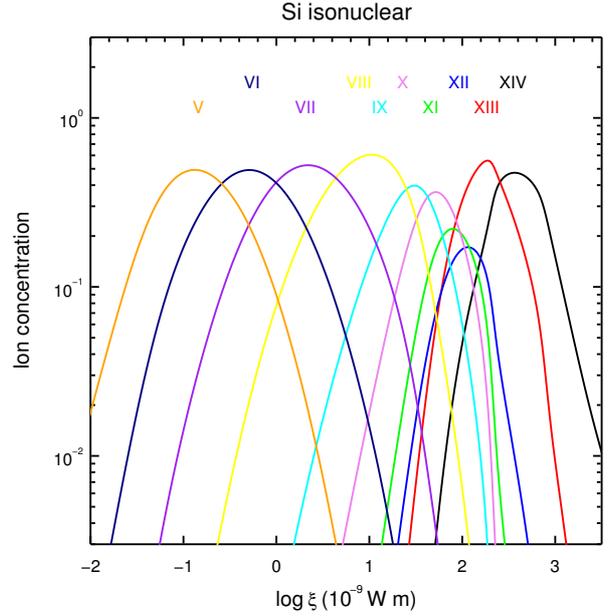}
\caption{Ion concentration of the Si isonuclear sequence (H-like to Ne-like) as a function of the ionization parameter (in units of $10^{-9}$~W~m, i.e., erg~${\rm s^{-1}}$~cm).}
\label{fig:icon_spex_dev_pion_14ins}
\end{figure}

When modeling a photoionized plasma with our new PION model in the SPEX code, its thermal equilibrium, ionization balance, and level population are calculated self-consistently with detailed atomic data of relevant collisional and radiative processes, such as collisional excitation and de-excitation \citep[FAC calculation,][]{gu08}, radiative recombination \citep{bad06, mao16}, inner shell ionization \citep{urd17}, etc. The thermal balance and instability curve are shown in Figs.~3--5 of \citet{meh16}. Ion concentrations are derived accordingly, and we show in Figure~\ref{fig:icon_spex_dev_pion_14ins} the ion concentrations of \ion{Si}{XIV} (H-like) to \ion{Si}{V} (Ne-like) as a function of the ionization parameter. The level population is also calculated simultaneously.

\section{Results}
\label{sct:04to06ies}
In Section~\ref{sct:04ies} (for Be-like ions) to Section~\ref{sct:06ies} (for C-like ions), we first show which metastable levels (Table~\ref{tbl:lev_idx}) can be significantly populated in a broad density range of $n_{\rm H}\in(10^{6},~10^{20})~{\rm m^{-3}}$. We note that different ionization parameters are used to maximize the ion concentration of different ions, e.g., $\log_{10} (\xi) = 1.90$ for \ion{Si}{XI} (Be-like), $\log_{10} (\xi) = 2.0$ for \ion{S}{XII} (B-like), and $\log_{10} (\xi) = 2.1$ for \ion{Ar}{XIII} (C-like). Second, we list the characteristic absorption lines from these metastable levels. Absorption lines from the ground level are listed together with density sensitive lines from the metastable levels, so that it is possible to tell whether the spectral resolution of a certain instrument is fine enough to distinguish these lines. If so, these lines can be used for density diagnostics. 

\begin{table*}
\caption{Level indices for the ground level (Level 1) and density sensitive metastable levels (Levels 2-5) from the Be-like to F-like isoelectronic sequences.}
\label{tbl:lev_idx}
\centering
\begin{tabular}{llllllllllll}
\hline\hline
\noalign{\smallskip}
Index & \multicolumn{2}{c}{1} &  \multicolumn{2}{c}{2} &  \multicolumn{2}{c}{3} & \multicolumn{2}{c}{4} & \multicolumn{2}{c}{5} \\
\noalign{\smallskip}
\hline\hline
\noalign{\smallskip}
Sequence & Conf. & $^{2S+1}L_{J}$ & Conf.  & $^{2S+1}L_{J}$  & Conf.  & $^{2S+1}L_{J}$  & Conf.  & $^{2S+1}L_{J}$ & Conf.  & $^{2S+1}L_{J}$  \\
\noalign{\smallskip}
\hline
\noalign{\smallskip}
Be-like & $2s^{2}$ & $^{1}S_{0}$ & $2s 2p$ & $^{3}P_{0}$ & $2s 2p$ & $^{3}P_{1}$ & $2s 2p$ & $^{3}P_{2}$  &  -- -- & -- -- \\
\noalign{\smallskip}
B-like & $2s^{2} 2p$ & $^{2}P_{1/2}$ & $2s^{2} 2p$ & $^{2}P_{3/2}$ &  -- -- & -- -- &  -- -- & -- -- &  -- -- & -- --  \\
\noalign{\smallskip}
C-like & $2s^{2} 2p^{2}$ & $^{3}P_{0}$ & $2s^{2} 2p^{2}$ & $^{3}P_{1}$ & $2s^{2} 2p^{2}$ & $^{3}P_{2}$ & $2s^{2} 2p^{2}$ & $^{1}D_{2}$ & $2s^{2} 2p^{2}$ & $^{1}S_{0}$  \\
\noalign{\smallskip}
N-like (Fe) & $2s^{2} 2p^{3}$ & $^{4}S_{3/2}$ & $2s^{2} 2p^{3}$ & $^{2}D_{3/2}$ & $2s^{2} 2p^{3}$ & $^{2}D_{5/2}$ & $2s^{2} 2p^{3}$ & $^{2}P_{1/2}$ & $2s^{2} 2p^{3}$ & $^{2}P_{3/2}$ \\
\noalign{\smallskip}
O-like (Fe) & $2s^{2} 2p^{4}$ & $^{3}P_{2}$ & $2s^{2} 2p^{4}$ & $^{3}P_{0}$ & $2s^{2} 2p^{4}$ & $^{3}P_{1}$ & $2s^{2} 2p^{4}$ & $^{1}D_{2}$ & $2s^{2} 2p^{4}$ & $^{1}S_{0}$   \\
\noalign{\smallskip}
F-like (Fe) & $2s^{2} 2p^{5}$ & $^{2}P_{3/2}$ & $2s^{2} 2p^{5}$ & $^{2}P_{1/2}$ &  -- -- & -- -- &  -- -- & -- -- &  -- -- & -- --   \\
\noalign{\smallskip}
 \hline
\end{tabular}
\tablefoot{``Conf.'' is short for electron configuration. $^{2S+1}L_{J}$ refers to the spectroscopic notations, where $S$ is the total spin quantum number ($2S + 1$ is the spin multiplicity), $L$ is the total orbital quantum number, and $J$ is the total angular momentum quantum number.}
\end{table*}

\subsection{Be-like}
\label{sct:04ies}
\begin{figure*}
\centering
\includegraphics[width=\hsize]{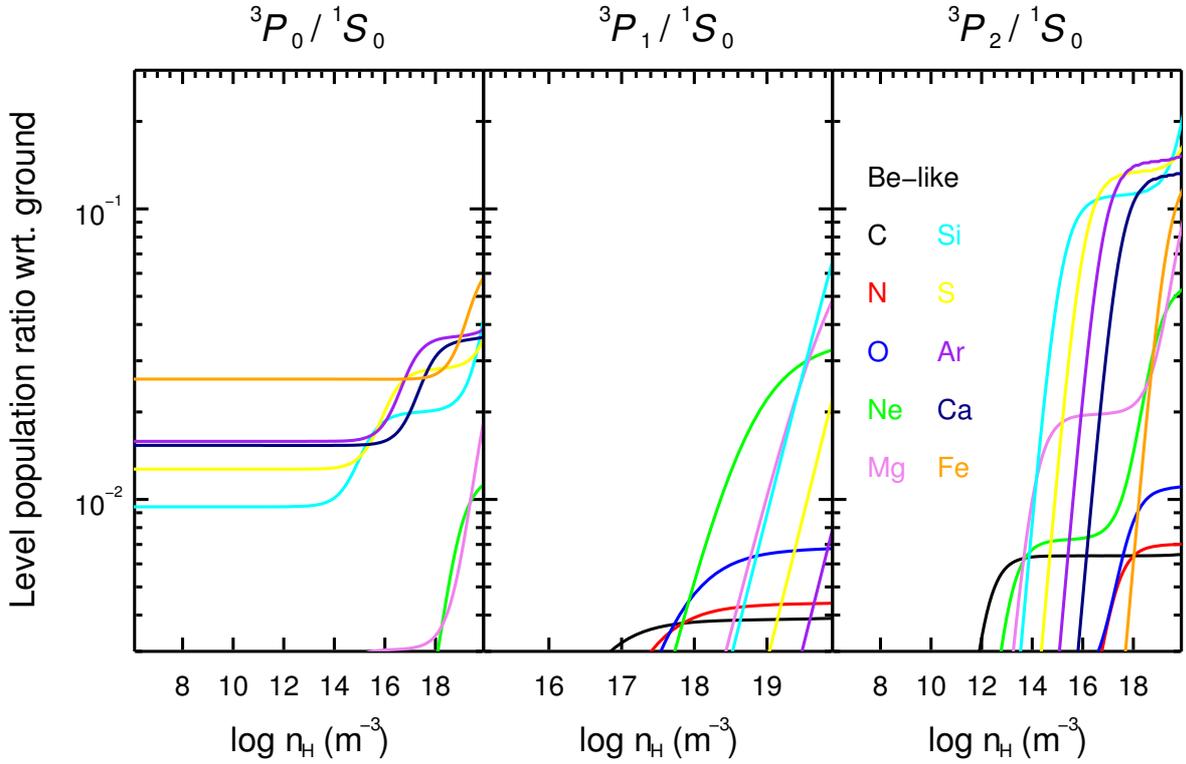}
\caption{Level population ratios as a function of plasma density (in the range of ${\rm 10^{6-20} m^{-3}}$ or ${\rm 10^{0-14} cm^{-3}}$) at ionization parameter of maximum ion concentration in the ionization balance. The configuration of the ground ($^1S_0$) and metastable levels ($^3P_{0-2}$) are listed in Table~\ref{tbl:lev_idx}.}
\label{fig:lpop_spex_dev_pion_4ies}
\end{figure*}

Figure~\ref{fig:lpop_spex_dev_pion_4ies} shows the ratios of  metastable to ground level populations as a function of plasma density. The lifetime of the first excited level $1s^2 2s 2p~(^3P_0)$ is rather long, so that even if the plasma density is rather low, the metastable level can still be populated up to a small percent of the ground level population. Accordingly, a plateau can be found in the $^3P_0/^3S_0$ level population ratio (left panel of Figure~\ref{fig:lpop_spex_dev_pion_4ies}). The third excited level $1s^2 2s 2p~(^3P_2)$ can be more easily populated at lower density  than the second excited level $1s^2 2s 2p~(^3P_1)$. The rest of the excited levels in Be-like ions are not significantly populated ($<0.01\%$ of the ground level population) in a photoionized plasma with density $n_{\rm H} \lesssim 10^{20}~{\rm m^{-3}}$.

Given that the metastable levels (Level 2-4) can be populated up to 20\% of the ground level (Level 1) population, we list in Table~\ref{tbl:cal_4ies} three sets of characteristic transitions ($n_j=2-2,~2-3,~{\rm and}~1-2$) for each level. The corresponding wavelengths ($\lambda$) and oscillator strengths ($f$) of these absorption features in the X-ray ($1-100$~\AA) and UV ($100-2000$~\AA) wavelength ranges are listed as well.

\begin{table*}
\caption{Characteristic absorption lines from the ground and the metastable levels in Be-like ions.}
\label{tbl:cal_4ies}
\centering
\small
\begin{tabular}{llllllllllll}
\hline\hline
\noalign{\smallskip}
Index & \multicolumn{2}{l}{1} &  \multicolumn{2}{l}{2} &  \multicolumn{2}{l}{3} & \multicolumn{2}{l}{4}   \\
\noalign{\smallskip}
\hline\hline
\noalign{\smallskip}
$n_j$ & Lower & Upper & Lower & Upper & Lower & Upper & Lower & Upper \\
\noalign{\smallskip}
\hline
\noalign{\smallskip}
$2-2$ & $2s^2~(^1S_{0})$ & $2s 2p~(^1P_{1})$ & $2s 2p~(^3P_{0})$ & $2p^2~(^3P_{1})$ & $2s 2p~(^3P_{1})$ & $2p^2~(^3P_{2})$ & $2s 2p~(^3P_{2})$ & $2p^2~(^3P_{2})$ \\
\noalign{\smallskip}
\hline
\noalign{\smallskip}
Ion & $\lambda$~(\AA) & $f$ & $\lambda$~(\AA) & $f$ & $\lambda$~(\AA) & $f$ & $\lambda$~(\AA) & $f$ \\
\noalign{\smallskip}
\hline
\noalign{\smallskip}
\ion{C}{III} & 977.012 & 0.76 & 1175.254 & 0.27 & 1174.923 & 0.11 & 1175.702 & 0.20 \\ 
\ion{N}{IV} & 765.141 & 0.61 & 922.511 & 0.22 & 921.986 & 0.09 & 923.212 & 0.17 \\ 
\ion{O}{V} & 629.727 & 0.50 & 759.436 & 0.19 & 758.671 & 0.08 & 760.440 & 0.14 \\ 
\ion{Ne}{VII} & 465.216 & 0.41 & 559.944 & 0.16 & 558.605 & 0.07 & 561.724 & 0.12 \\ 
\ion{Mg}{IX} & 368.068 & 0.31 & 441.196 & 0.12 & 439.173 & 0.05 & 443.969 & 0.09 \\ 
\ion{Si}{XI} & 303.323 & 0.27 & 361.410 & 0.10 & 358.650 & 0.04 & 365.431 & 0.08 \\ 
\ion{S}{XIII} & 256.683 & 0.24 & 303.382 & 0.09 & 299.954 & 0.04 & 308.950 & 0.07 \\ 
\ion{Ar}{XV} & 221.133 & 0.21 & 258.764 & 0.08 & 254.827 & 0.04 & 266.239 & 0.06 \\ 
\ion{Ca}{XVII} & 192.818 & 0.19 & 223.018 & 0.08 & 218.821 & 0.03 & 232.827 & 0.05 \\ 
\ion{Fe}{XXIII} & 132.906 & 0.15 & 147.270 & 0.06 & 144.389 & 0.03 & 166.689 & 0.03 \\ 
\noalign{\smallskip}
\hline
\noalign{\smallskip}
$n_j$ & Lower & Upper & Lower & Upper & Lower & Upper & Lower & Upper \\
\noalign{\smallskip}
\hline
\noalign{\smallskip}
$2-3$ & $2s^2~(^1S_{0})$ & $2s 3p~(^1P_{1})$ & $2s 2p~(^3P_{0})$ & $2s 3d~(^3D_{1})$ & $2s 2p~(^3P_{1})$ & $2s 3d~(^3D_{2})$ & $2s 2p~(^3P_{2})$ & $2s 3d~(^3D_{3})$ \\
\noalign{\smallskip}
\hline
\noalign{\smallskip}
Ion & $\lambda$~(\AA) & $f$ & $\lambda$~(\AA) & $f$ & $\lambda$~(\AA) & $f$ & $\lambda$~(\AA) & $f$ \\
\noalign{\smallskip}
\hline
\noalign{\smallskip}
\ion{C}{III} & 386.200 & 0.23 & 459.463 & 0.56 & 459.510 & 0.42 & 459.624 & 0.47 \\ 
\ion{N}{IV} & 247.203 & 0.33 & 283.415 & 0.61 & 283.463 & 0.46 & 283.572 & 0.51 \\ 
\ion{O}{V} & 172.168 & 0.38 & 192.749 & 0.63 & 192.795 & 0.47 & 192.902 & 0.53 \\ 
\ion{Ne}{VII} & 97.495 & 0.47 & 106.040 & 0.68 & 106.085 & 0.51 & 106.189 & 0.57 \\ 
\ion{Mg}{IX} & 62.751 & 0.53 & 67.089 & 0.72 & 67.134 & 0.54 & 67.239 & 0.61 \\ 
\ion{Si}{XI} & 43.763 & 0.46 & 46.362 & 0.73 & 46.298 & 0.54 & 46.399 & 0.61 \\ 
\ion{S}{XIII} & 32.242 & 0.38 & 33.806 & 0.74 & 33.852 & 0.55 & 33.951 & 0.61 \\ 
\ion{Ar}{XV} & 24.759 & 0.32 & 25.808 & 0.75 & 25.850 & 0.56 & 25.953 & 0.62 \\ 
\ion{Ca}{XVII} & 19.558 & 0.36 & 20.310 & 0.78 & 20.339 & 0.56 & 20.437 & 0.62 \\ 
\ion{Fe}{XXIII} & 10.980 & 0.41 & 11.298 & 0.74 & 11.325 & 0.54 & 11.441 & 0.61 \\ 
\noalign{\smallskip}
\hline
\noalign{\smallskip}
$n_j$ & Lower & Upper & Lower & Upper~($\dagger$) & Lower & Upper~($\dagger$) & Lower & Upper~($\dagger$) \\
\noalign{\smallskip}
\hline
\noalign{\smallskip}
$1-2$ & $2s^2~(^1S_{0})$ & $1s 2s^2 2p~(^1P_{1})$ & $2s 2p~(^3P_{0})$ & $1s 2s 2p^2~(^3P_1)$ & $2s 2p~(^3P_{1})$ & $1s 2s 2p^2~(^3P_2)$ & $2s 2p~(^3P_{2})$ & $1s 2s 2p^2~(^3P_2)$ \\
\noalign{\smallskip}
\hline
\noalign{\smallskip}
Ion & $\lambda$~(\AA) & $f$ & $\lambda$~(\AA) & $f$ & $\lambda$~(\AA) & $f$ & $\lambda$~(\AA) & $f$ \\
\noalign{\smallskip}
\hline
\noalign{\smallskip}
\ion{C}{III} & 42.165 & 0.62 & 42.385 & 0.31 & 42.384 & 0.13 & 42.385 & 0.23 \\ 
\ion{N}{IV} & 29.941 & 0.65 & 30.088 & 0.33 & 30.086 & 0.14 & 30.088 & 0.24 \\ 
\ion{O}{V} & 22.360 & 0.61 & 22.474 & 0.30 & 22.472 & 0.13 & 22.474 & 0.22 \\ 
\ion{Ne}{VII} & 13.820 & 0.66 & 13.878 & 0.35 & 13.877 & 0.16 & 13.879 & 0.22 \\ 
\ion{Mg}{IX} & 9.378 & 0.68 & 9.410 & 0.44 & 9.413 & 0.26~($^3D_2$) & 9.415 & 0.15~($^3D_3$) \\ 
\ion{Si}{XI} & 6.776 & 0.70 & 6.798 & 0.52~($^3D_1$) & 6.799 & 0.27~($^3D_2$) & 6.797 & 0.27 \\ 
\ion{S}{XIII} & 5.123 & 0.71 & 5.137 & 0.56~($^3D_1$) & 5.138 & 0.27~($^3D_2$) & 5.136 & 0.28 \\ 
\ion{Ar}{XV} & 4.007 & 0.72 & 4.017 & 0.57~($^3D_1$) & 4.018 & 0.27~($^3D_2$) & 4.016 & 0.29 \\ 
\ion{Ca}{XVII} & 3.222 & 0.72 & 3.228 & 0.59~($^3D_1$) & 3.228 & 0.28~($^3D_2$) & 3.228 & 0.29 \\ 
\ion{Fe}{XXIII} & 1.870 & 0.69 & 1.873 & 0.61 & 1.874 & 0.29~($^3D_2$) & 1.874 & 0.26 \\ 
\hline
\end{tabular}
\tablefoot{For the $1s-2p$ transitions (denoted $n_j=1-2$) from the metastable (lower) levels, the upper levels (marked  $\dagger$) vary ($^3P_{1-2}$ or $^3D_{1-3}$) for different elements (exceptions are marked following the $f$-values). }
\end{table*}

Among the three transitions of each level in the same ion, the inner shell $1s-2p$ transition (denoted as $n_j=1-2$) always yields the shortest wavelength, while the $2s-2p$ transition ($n_j=2-2$) yields the longest wavelength. There are of course more transitions than we listed in Table~\ref{tbl:cal_4ies}. For instance, there are in total six $1s^2 2s 2p~(^3P)$ -- $1s^2 2p^2~(^3P)$ metastable transitions around $\lambda\sim1175$~\AA\ for \ion{C}{III} that have been successfully used for density diagnostics in AGN ionized outflows \citep[e.g.,][]{ara15}. Transitions with higher $f$-values are listed in Table~\ref{tbl:cal_4ies} for simplicity. For the six \ion{C}{III} $\lambda\sim1175$~\AA\ lines, the three tabulated lines have $f>0.1$, while the other three lines have slightly lower oscillator strength values. 

For all three sets of transitions, the metastable to ground separation ($\Delta \lambda=|\lambda_{2, 3, 4} - \lambda_{1}|$) increases with increasing wavelength of the lines. For the inner shell transitions, the separations are rather small with $\Delta \lambda (n_j=1-2) \lesssim0.14$~\AA. For the other two sets of transitions, the separations are relatively large, ranging from $\sim0.3$~\AA\ (for $\lambda\sim10$~\AA) to $\sim200$~\AA\ (for $\lambda\sim1000$~\AA), which are larger than the spectral resolution of current X-ray grating spectrometers. The metastable lines themselves, on the other hand, are closer to each other ($\lesssim 0.2$~\AA\ for lines in the X-ray band and $\lesssim 3$~\AA\ for lines in the UV band). This is simply due to the fact that the three metastable levels are the fine structure splitting of the same $^3P$ term.

\subsection{B-like}
\label{sct:05ies}
\begin{figure}
\centering
\includegraphics[width=\hsize, trim={0.3cm 0.3cm 0.3cm 0.3cm}, clip]{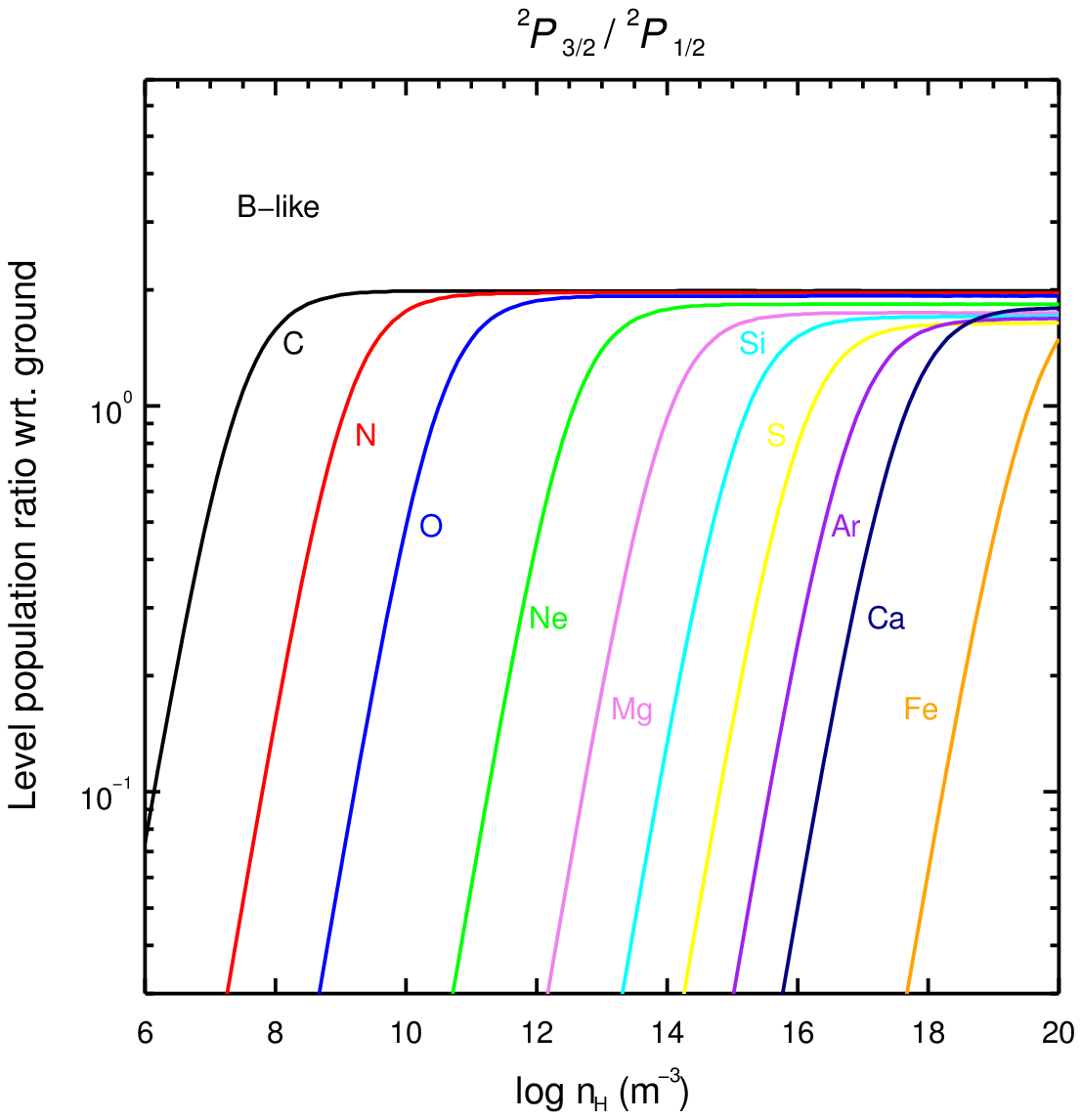}
\caption{Similar to Figure~\ref{fig:lpop_spex_dev_pion_4ies}, but for B-like ions.}
\label{fig:lpop_spex_dev_pion_5ies}
\end{figure}

For B-like ions, only the first excited level $2s^2~2p~(^2P_{3/2})$ can be significantly populated up to a factor of two above the ground level population, as shown in Figure~\ref{fig:lpop_spex_dev_pion_5ies}. None of the rest of the excited levels is significantly populated ($<0.01\%$ of the ground level population) in a photoionized plasma with density $n_{\rm H} \lesssim 10^{20}~{\rm m^{-3}}$. At a sufficiently high density, the ratio of metastable to ground level population  follows the Boltzmann distribution (meanwhile the plasma is partially in local temperature equilibrium). 

Characteristic absorption lines from the ground (Level 1) and metastable level (Level 2) in B-like ions are listed in Table~\ref{tbl:cal_5ies}. Similar to the Be-like isoelectronic sequence (Section~\ref{sct:04ies}), among the three transitions for the same ion, the inner shell transition ($n_j=1-2$) has the shortest wavelength and the $n_j=2-2$ transition has the longest wavelength. The metastable to ground separation ($\Delta \lambda$) is negligible for the inner shell transition (${n_j=1-2}$). For the other two sets of transitions, the separations between the ground and metastable lines are $\Delta \lambda({n_j=2-3}) \sim 0.15-0.30$~\AA\ and $\Delta \lambda({n_j=2-2}) \sim 0.2-2.7$~\AA, respectively, which are larger than the spectral resolution of current X-ray grating spectrometers \citep[see, e.g., Table~1 in][]{kaa17a}. 
\begin{table*}
\caption{Characteristic absorption lines from the ground and metastable level in B-like ions.}
\label{tbl:cal_5ies}
\centering
\begin{tabular}{llllllllllll}
\hline\hline
\noalign{\smallskip}
Index & \multicolumn{2}{l}{1} &  \multicolumn{2}{l}{2}   \\
\noalign{\smallskip}
\hline\hline
\noalign{\smallskip}
$n_j$ & Lower & Upper~($\dagger$) & Lower & Upper \\
\noalign{\smallskip}
\hline
\noalign{\smallskip}
$2-2$ & $2s^2 2p~(^2P_{1/2})$ & $2s 2p^2~(^2P_{1/2})$ & $2s^2 2p~(^2P_{3/2})$ & $2s 2p^2~(^2P_{3/2})$ \\
\noalign{\smallskip}
\hline
\noalign{\smallskip}
Ion & $\lambda$~(\AA) & $f$ & $\lambda$~(\AA) & $f$ \\
\noalign{\smallskip}
\hline
\noalign{\smallskip}
\ion{C}{II} & 903.962 & 0.33 & 904.142 & 0.42 \\ 
\ion{N}{III} & 685.515 & 0.28 & 685.818 & 0.36 \\ 
\ion{O}{IV} & 554.076 & 0.23 & 554.514 & 0.29 \\ 
\ion{Ne}{VI} & 401.146 & 0.17 & 401.941 & 0.22 \\ 
\ion{Mg}{VIII} & 313.754 & 0.13 & 315.039 & 0.18 \\ 
\ion{Si}{X} & 256.384 & 0.09 & 258.372 & 0.15 \\ 
\ion{S}{XII} & 227.490 & 0.07~($^2S_{1/2}$) & 218.200 & 0.13 \\ 
\ion{Ar}{XIV} & 194.401 & 0.08~($^2S_{1/2}$) & 187.962 & 0.12 \\ 
\ion{Ca}{XVI} & 168.868 & 0.08~($^2S_{1/2}$) & 164.165 & 0.11 \\ 
\ion{Fe}{XXII} & 117.144 & 0.08 & 114.409 & 0.09 \\ 
\noalign{\smallskip}
\hline
\noalign{\smallskip}
$n_j$ & Lower & Upper & Lower & Upper \\
\noalign{\smallskip}
\hline
\noalign{\smallskip}
$2-3$ & $2s^2 2p~(^2P_{1/2})$ & $2s^2 3d~(^2D_{3/2})$ & $2s^2 2p~(^2P_{3/2})$ & $2s^2 3d~(^2D_{5/2})$ \\
\noalign{\smallskip}
\hline
\noalign{\smallskip}
Ion & $\lambda$~(\AA) & $f$ & $\lambda$~(\AA) & $f$ \\
\noalign{\smallskip}
\hline
\noalign{\smallskip}
\ion{C}{II} & 687.053 & 0.33 & 687.346 & 0.30 \\ 
\ion{N}{III} & 374.198 & 0.44 & 374.434 & 0.39 \\ 
\ion{O}{IV} & 238.360 & 0.50 & 238.570 & 0.45 \\ 
\ion{Ne}{VI} & 122.516 & 0.56 & 122.701 & 0.50 \\ 
\ion{Mg}{VIII} & 74.858 & 0.60 & 75.034 & 0.54 \\ 
\ion{Si}{X} & 50.524 & 0.62 & 50.691 & 0.56 \\ 
\ion{S}{XII} & 36.399 & 0.63 & 36.564 & 0.57 \\ 
\ion{Ar}{XIV} & 27.469 & 0.64 & 27.631 & 0.58 \\ 
\ion{Ca}{XVI} & 21.451 & 0.65 & 21.609 & 0.58 \\ 
\ion{Fe}{XXII} & 11.767 & 0.67 & 11.921 & 0.59 \\ 
\noalign{\smallskip}
\hline
\noalign{\smallskip}
$n_j$ & Lower & Upper~($\dagger$) & Lower & Upper \\
\noalign{\smallskip}
\hline
\noalign{\smallskip}
$1-2$ & $2s^2 2p~(^2P_{1/2})$ & $1s 2s^2 2p^2~(^2P_{3/2})$ & $2s^2 2p~(^2P_{3/2})$ & $1s 2s^2 2p^2~(^2P_{3/2})$  \\
\noalign{\smallskip}
\hline
\noalign{\smallskip}
Ion & $\lambda$~(\AA) & $f$ & $\lambda$~(\AA) & $f$ \\
\noalign{\smallskip}
\hline
\noalign{\smallskip}
\ion{C}{II} & 43.050 & 0.13 & 43.049 & 0.17 \\ 
\ion{N}{III} & 30.533 & 0.16 & 30.532 & 0.20 \\ 
\ion{O}{IV} & 22.762 & 0.18 & 22.761 & 0.22 \\ 
\ion{Ne}{VI} & 14.030 & 0.20 & 14.030 & 0.25 \\ 
\ion{Mg}{VIII} & 9.500 & 0.22 & 9.500 & 0.28 \\ 
\ion{Si}{X} & 6.853 & 0.23 & 6.853 & 0.29 \\ 
\ion{S}{XII} & 5.179 & 0.25~($^2D_{3/2}$) & 5.174 & 0.30 \\ 
\ion{Ar}{XIV} & 4.048 & 0.27~($^2D_{3/2}$) & 4.046 & 0.31 \\ 
\ion{Ca}{XVI} & 3.249 & 0.29~($^2D_{3/2}$) & 3.247 & 0.31 \\ 
\ion{Fe}{XXII} & 1.882 & 0.32~($^2D_{3/2}$) & 1.882 & 0.30 \\ 
\hline
\end{tabular}
\tablefoot{For the $2s-2p~(n_j=2-2)$ and $1s-2p~(n_j=1-2)$ transitions from the ground (lower) level, the upper levels (marked  $\dagger$) vary  for different elements (exceptions are marked following the $f$-values).}
\end{table*}

\subsection{C-like}
\label{sct:06ies}
\begin{figure*}
\centering
\includegraphics[width=\hsize]{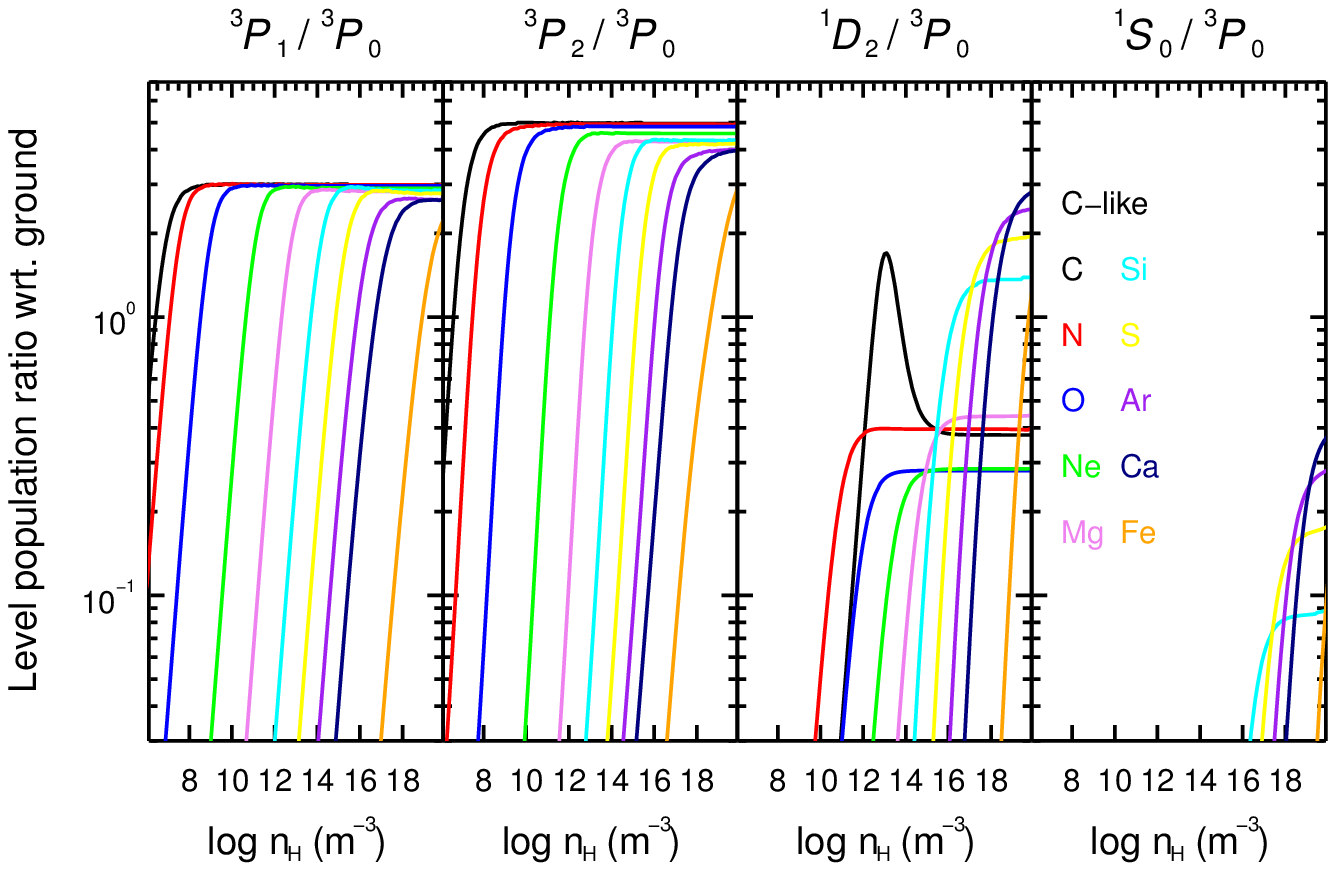}
\caption{Similar to Figure~\ref{fig:lpop_spex_dev_pion_4ies}, but for C-like ions..}
\label{fig:lpop_spex_dev_pion_6ies}
\end{figure*}

For C-like ions, the first two excited levels $2s^2~2p^2~(^3P_1~{\rm and}~^3P_2)$ can be significantly populated up to a factor of a few of the ground level population, as shown in Figure~\ref{fig:lpop_spex_dev_pion_6ies}. The next two excited levels $2s^2~2p^2~(^1D_2~\rm and~^1S_0)$ can also be populated up to tens of percent of the ground level population. The rest of the excited levels are not significantly populated ($<0.01\%$ of the ground level population) in a photoionized plasma with density $n_{\rm H} \lesssim 10^{20}~{\rm m^{-3}}$.

Since the metastable level $2s^2 2p^2~(^1S_{0})$ is only significantly populated at rather high density (the right  panel in Figure~\ref{fig:lpop_spex_dev_pion_6ies}), for simplicity only transitions from the ground (Level 1) and the first three metastable levels (Level $2-4$) are listed in Table~\ref{tbl:cal_6ies}. Again, the inner shell transition ($n_j=1-2$) corresponds to the line with the shortest wavelength and the $n_j=2-2$ transition corresponds to the line with the longest wavelength. For the inner shell transition, the ground to metastable separation ($\Delta \lambda$) is negligible. For the other two transitions ($n_j=2-2~{\rm and}~2-3$) that we listed here, it is easier (with $\Delta \lambda\gtrsim 0.1$~\AA) to distinguish lines from the ground $2s^2~2p^2~(^3P_{0})$ and lines from the third excited level $2s^2~2p^2~(^1D_{2})$. Lines from the ground and the first two excited levels are close to each other since the lower levels are the fine structure splitting of the same term $^3P$. Between the $n_j=2-3$ and $n_j=2-2$ transitions, it is more difficult to distinguish lines from the ground and the first two excited levels for the former, e.g., \ion{C}{I} lines around $\lambda \sim1277$~\AA, \ion{N}{II} lines around $\lambda \sim533$~\AA.

\begin{table*}
\caption{Characteristic absorption lines from the ground and the metastable levels in C-like ions.}
\label{tbl:cal_6ies}
\centering
\small
\begin{tabular}{llllllllllll}
\hline\hline
\noalign{\smallskip}
Index & \multicolumn{2}{l}{1} &  \multicolumn{2}{l}{2} &  \multicolumn{2}{l}{3} & \multicolumn{2}{l}{4}  \\
\noalign{\smallskip}
\hline\hline
\noalign{\smallskip}
$n_j$ & Lower & Upper~($\dagger$) & Lower & Upper~($\dagger$) & Lower & Upper~($\dagger$) & Lower & Upper \\
\noalign{\smallskip}
\hline
\noalign{\smallskip}
$2-2$ & $2s^2 2p^2~(^3P_{0})$ & $2s 2p^3~(^3S_1)$ & $2s^2 2p^2~(^3P_{1})$ & $2s 2p^3~(^3S_1)$ & $2s^2 2p^2~(^3P_{2})$ & $2s 2p^3~(^3S_1)$ & $2s^2 2p^2~(^1D_{2})$ & $2s 2p^3~(^1D_{2})$  \\
\noalign{\smallskip}
\hline
\noalign{\smallskip}
Ion & $\lambda$~(\AA) & $f$ & $\lambda$~(\AA) & $f$ & $\lambda$~(\AA) & $f$ & $\lambda$~(\AA) & $f$  \\
\noalign{\smallskip}
\hline
\noalign{\smallskip}
\ion{C}{I} & 1560.310 & 0.18~$(^3D_1)$ & 1560.683 & 0.14~$(^3D_2)$ & 1561.439 & 0.15~$(^3D_3)$ & 1021.853 & 0.61 \\ 
\ion{N}{II} & 644.634 & 0.25 & 644.837 & 0.25 & 645.179 & 0.25 & 775.966 & 0.36 \\ 
\ion{O}{III} & 507.389 & 0.20 & 507.680 & 0.20 & 508.178 & 0.20 & 599.590 & 0.34 \\ 
\ion{Ne}{V} & 357.947 & 0.15 & 358.475 & 0.15 & 359.374 & 0.15 & 416.212 & 0.27 \\ 
\ion{Mg}{VII} & 276.154 & 0.12 & 277.001 & 0.12 & 278.402 & 0.12 & 319.027 & 0.22 \\ 
\ion{Si}{IX} & 223.743 & 0.10 & 225.024 & 0.10 & 227.000 & 0.10 & 258.080 & 0.19 \\ 
\ion{S}{XI} & 186.839 & 0.08 & 188.675 & 0.08 & 191.266 & 0.09 & 215.968 & 0.16 \\ 
\ion{Ar}{XIII} & 236.268 & 0.08~$(^3D_1)$ & 161.610 & 0.07 & 164.802 & 0.08 & 184.899 & 0.14 \\ 
\ion{Ca}{XV} & 200.976 & 0.08~$(^3D_1)$ & 140.583 & 0.06 & 144.309 & 0.07 & 161.018 & 0.12 \\ 
\ion{Fe}{XXI} & 128.752 & 0.09~$(^3D_1)$ & 142.149 & 0.05~$(^3D_2)$ & 102.216 & 0.06 & 113.292 & 0.09 \\ 
\noalign{\smallskip}
\hline
\noalign{\smallskip}
$n_j$ & Lower & Upper & Lower & Upper & Lower & Upper & Lower & Upper \\
\noalign{\smallskip}
\hline
\noalign{\smallskip}
$2-3$ & $2s^2 2p^2~(^3P_{0})$ & $2s^2 2p 3d~(^3D_{1})$ & $2s^2 2p^2~(^3P_{1})$ & $2s^2 2p 3d~(^3D_{2})$ & $2s^2 2p^2~(^3P_{2})$ & $2s^2 2p 3d~(^3D_{3})$ & $2s^2 2p^2~(^1D_{2})$ & $2s^2 2p 3d~(^1F_{3})$  \\
\noalign{\smallskip}
\hline
\noalign{\smallskip}
Ion & $\lambda$~(\AA) & $f$ & $\lambda$~(\AA) & $f$ & $\lambda$~(\AA) & $f$ & $\lambda$~(\AA) & $f$  \\
\noalign{\smallskip}
\hline
\noalign{\smallskip}
\ion{C}{I} & 1277.246 & 0.11 & 1277.283 & 0.08 & 1277.550 & 0.10 & 1463.337 & 0.10 \\ 
\ion{N}{II} & 533.511 & 0.33 & 533.582 & 0.25 & 533.729 & 0.27 & 574.650 & 0.29 \\ 
\ion{O}{III} & 305.596 & 0.54 & 305.656 & 0.37 & 305.768 & 0.41 & 320.978 & 0.46 \\ 
\ion{Ne}{V} & 143.220 & 0.76 & 143.265 & 0.57 & 143.345 & 0.58 & 147.137 & 0.69 \\ 
\ion{Mg}{VII} & 83.910 & 0.93 & 83.959 & 0.68 & 84.025 & 0.66 & 85.407 & 0.83 \\ 
\ion{Si}{IX} & 55.305 & 1.05 & 55.356 & 0.74 & 55.401 & 0.71 & 56.027 & 0.91 \\ 
\ion{S}{XI} & 39.240 & 1.13 & 39.300 & 0.74 & 39.323 & 0.73 & 39.648 & 0.97 \\ 
\ion{Ar}{XIII} & 29.318 & 1.19 & 29.252 & 0.68 & 29.348 & 0.73 & 29.549 & 1.00 \\ 
\ion{Ca}{XV} & 22.730 & 1.23 & 22.758 & 0.57 & 22.777 & 0.72 & 22.902 & 1.00 \\ 
\ion{Fe}{XXI} & 12.292 & 1.24 & 12.433 & 0.43 & 12.331 & 0.59 & 12.411 & 0.94 \\
\noalign{\smallskip}
\hline
\noalign{\smallskip}
$n_j$ & Lower & Upper~($\dagger$) & Lower & Upper~($\dagger$) & Lower & Upper~($\dagger$) & Lower & Upper~($\dagger$) \\
\noalign{\smallskip}
\hline
\noalign{\smallskip}
$1-2$ & $2s^2 2p^2~(^3P_{0})$ & $1s 2s^2 2p^3~(^3D_1)$ & $2s^2 2p^2~(^3P_{1})$ & $1s 2s^2 2p^3~(^3P_1)$ & $2s^2 2p^2~(^3P_{2})$ & $1s 2s^2 2p^3~(^3D_3)$ & $2s^2 2p^2~(^1D_{2})$ & $1s 2s^2 2p^3~(^1D_2)$  \\
\noalign{\smallskip}
\hline
\noalign{\smallskip}
Ion & $\lambda$~(\AA) & $f$ & $\lambda$~(\AA) & $f$ & $\lambda$~(\AA) & $f$ & $\lambda$~(\AA) & $f$  \\
\noalign{\smallskip}
\hline
\noalign{\smallskip}
\ion{C}{I} & 43.567 & 0.09 & 43.568 & 0.08~($^3D_1$) & 43.568 & 0.08~($^3D_1$) & 43.492 & 0.17 \\ 
\ion{N}{II} & 30.963 & 0.11 & 30.944 & 0.10~($^3S_1$) & 30.945 & 0.10~($^3S_1$) & 30.914 & 0.21 \\ 
\ion{O}{III} & 23.056 & 0.13~($^3S_1$) & 23.057 & 0.11~($^3S_1$) & 23.058 & 0.11~($^3S_1$) & 23.061 & 0.24 \\ 
\ion{Ne}{V} & 14.234 & 0.16 & 14.205 & 0.14~($^3S_1$) & 14.237 & 0.12 & 14.211 & 0.28 \\ 
\ion{Mg}{VII} & 9.604 & 0.19~($^3S_1$) & 9.605 & 0.16~($^3S_1$) & 9.628 & 0.14 & 9.611 & 0.30 \\ 
\ion{Si}{IX} & 6.912 & 0.21~($^3P_1$) & 6.914 & 0.18 & 6.929 & 0.14 & 6.916 & 0.32 \\ 
\ion{S}{XI} & 5.225 & 0.24 & 5.218 & 0.19 & 5.229 & 0.15 & 5.220 & 0.33 \\ 
\ion{Ar}{XIII} & 4.086 & 0.27 & 4.081 & 0.20 & 4.088 & 0.15 & 4.081 & 0.33 \\ 
\ion{Ca}{XV} & 3.277 & 0.32 & 3.274 & 0.22 & 3.280 & 0.15 & 3.275 & 0.32 \\ 
\ion{Fe}{XXI} & 1.895 & 0.44 & 1.894 & 0.22~($^3S_1$) & 1.894 & 0.17~($^1D_2$) & 1.894 & 0.19~($^3P_2$) \\ 
\hline
\end{tabular}
\tablefoot{For the $2s-2p~(n_j=2-2)$ and $1s-2p~(n_j=1-2)$ transitions from the ground and metastable (lower) levels, the upper levels (marked  $\dagger$) vary for different element (exceptions are marked following the $f$-values). }
\end{table*}

\subsection{Summary}
\label{sct:sum}
For Be-like ions, the first three excited levels $2s 2p~(^3P_{0-2})$ can be significantly populated ($\gtrsim 1\%$ of the ground level population) when $n_{\rm H} \gtrsim 10^{14}~{\rm m^{-3}}$. Moreover, the third excited level  $2s 2p~(^3P_{2})$ can be populated more easily at such a high density. For B-like ions, merely the first excited level $2s^2 2p~(^2P_{3/2})$ can be used as a density probe. For C-like ions, the metastable levels $2s^2 2p^2~(^3P_1~{\rm and}~^3P_2)$ can be more easily populated than $2s^2 2p^2~(^1D_{2}~{\rm and}~^1S_{0})$. On the other hand, the transition from the $2s^2 2p^2~(^1D_{2})$ level can be more easily distinguished from the transition from the ground level $2s^2 2p^2~(^3P_{0})$. 

The metastable levels in Be-like ions are less populated than B-like and C-like ions at the same density. For all three isoelectronic sequences, it is rather difficult to distinguish inner shell transitions from the ground and metastable levels. 

The wavelengths and oscillator strengths can be obtained from the SPEX atomic code and table (SPEXACT) v3.04. We cross-checked the wavelengths of the lines in Tables~\ref{tbl:cal_4ies} to \ref{tbl:cal_6ies} with those from NIST v5.3 \citep{kra15} when available.

\section{Discussion}
\label{sct:dis}
\subsection{Ionization parameter dependence}
\label{sct:xidep}
In Section~\ref{sct:04to06ies}, for each ion, the level population is calculated given a certain ionization parameter. Here we discuss the level population dependence on both ionization parameter and density. Since the majority of the $n_j=2-2$ and $n_j=2-3$ transitions from ions in the same isoelectronic sequence share the same lower and upper levels (Table~\ref{tbl:cal_4ies} to \ref{tbl:cal_6ies}), we take these transitions from Be-like \ion{Si}{XI} to C-like \ion{Si}{IX} for this exercise. The ionization dependence can be applied to the rest of the ions in the same sequence. We note that inner shell transitions ($n_j=1-2$) are excluded since it is difficult to distinguish lines from ground and metastable levels in any case (Section~\ref{sct:sum}). We choose three different ionization parameters for each ion, corresponding to the maximum ion concentration and $\sim90\%$ of the maximum ion concentration. 

\begin{figure*}[!h]
\centering
\includegraphics[width=\hsize]{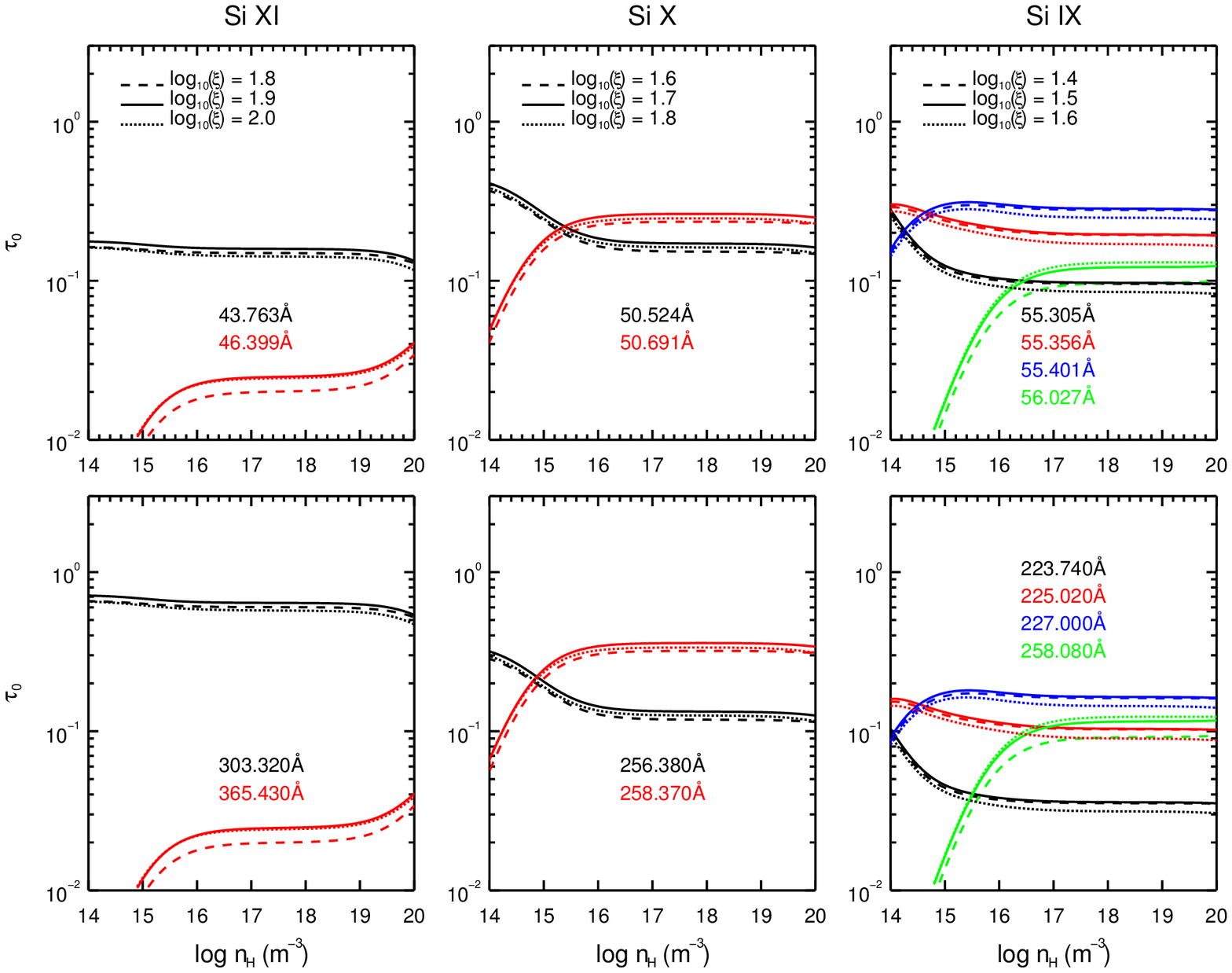}
\caption{Optical depth at the line center ($\tau_0$) for characteristic lines from ground and metastable levels in Be-like \ion{Si}{XI} to C-like \ion{Si}{IX}. The assumptions (SED, geometry, column density, turbulence, etc.) of the calculation are described in Section~\ref{sct:mo}.}
\label{fig:tau0_spex_dev_pion_14ins}
\end{figure*}

As shown in Figure~\ref{fig:tau0_spex_dev_pion_14ins}, the optical depth at the line center  ($\tau_0 \propto N_i~f$) varies with ionization parameter and density. Similarly, the metastable to ground equivalent width ratios ($EW_{\rm meta} / EW_{\rm ground}$) involving the $2s 2p~(^3P_2)$ metastable level in Be-like ions and the $2s^2 2p^2~(^1D_2)$ metastable level in C-like ions are sensitive to both ionization parameter and density of the plasma. Nonetheless, the ratios of metastable to ground equivalent width  involving the $2s^2 2p~(^2P_{3/2})$ metastable level in the B-like sequence and the $2s^2 2p^2~(^3P_1~{\rm and}~^3P_2)$ metastable levels in the C-like sequence are only sensitive to the density of the plasma, which makes them ideal density probes.

\begin{figure*}[!h]
\centering
\includegraphics[width=\hsize]{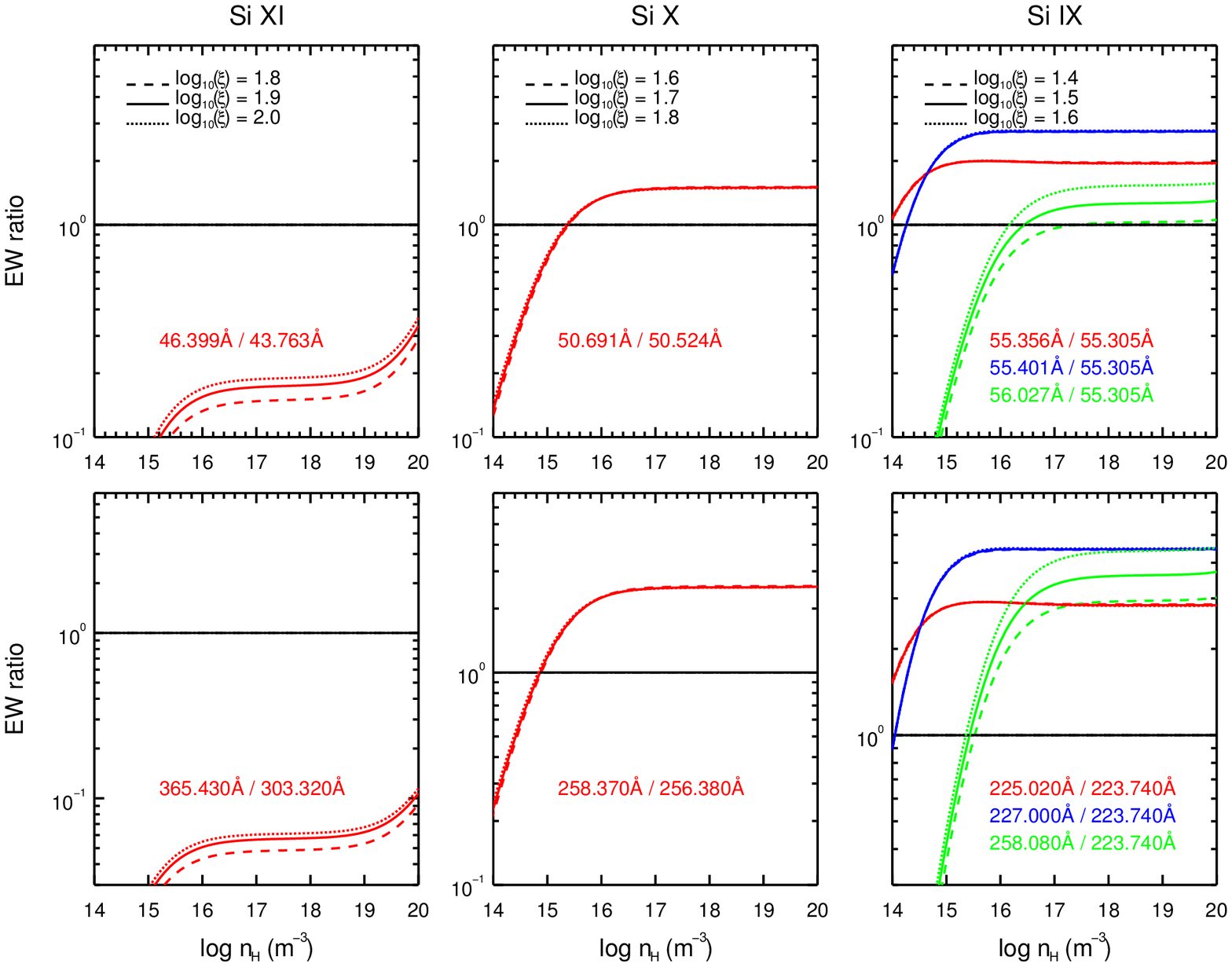}
\caption{Equivalent width (EW) ratios for characteristic lines from ground and metastable levels in Be-like \ion{Si}{XI} to C-like \ion{Si}{IX}. The assumptions (SED, geometry, column density, turbulence, etc.) of the calculation are described in Section~\ref{sct:mo}. For those lines with non-negligible optical depth $(\tau_0\gtrsim1$), the exact EWs depend on the line broadening profiles.}
\label{fig:ewrt_spex_dev_pion_14ins}
\end{figure*}

We caution that adopting a different ionizing SED yields a different ionization balance for the plasma. That is to say, the exact value of the ionization parameters where the ion concentration reaches a maximum or 90\% of the maximum in Figure~\ref{fig:ewrt_spex_dev_pion_14ins} differs for different SEDs (Table~\ref{tbl:xi_sed}). The SED adopted here is representative of a typical Seyfert 1 galaxy. Of course, the exact values of $\tau_0$ and EW also depend on the hydrogen column density ($N_{\rm H}$) and turbulence velocity ($v_{\rm b}$) of the ionized outflow. 
\begin{table}
\caption{Values of $\log_{10} (\xi)$ where Be-like \ion{Si}{XI} to C-like \ion{Si}{IX} reach the maximum ion concentration in the ionization balance adopting different ionizing SED. }
\label{tbl:xi_sed}
\centering
\begin{tabular}{lcccccc}
\hline\hline
\noalign{\smallskip}
SED & AGN1 & AGN2 & PL  \\
\hline
\noalign{\smallskip}
\ion{Si}{XI} & 1.90 & 2.03 & 1.90 \\
\ion{Si}{X} & 1.72 & 1.94 & 1.71 \\
\ion{Si}{IX} & 1.48 & 1.82 & 1.48 \\
\noalign{\smallskip}
\hline
\end{tabular}
\tablefoot{The three SEDs are adapted from \citet{meh16}.  AGN1 and AGN2 refer to the SEDs of an archetypal Seyfert 1 galaxy NGC\,5548 in a normal and obscured state. PL refers to a SED that follows the power law with $\Gamma=2$. }
\end{table}

\subsection{Domain of density and ionization parameter diagnostics}
\label{sct:domain}
\begin{figure}[!h]
\centering
\includegraphics[width=\hsize]{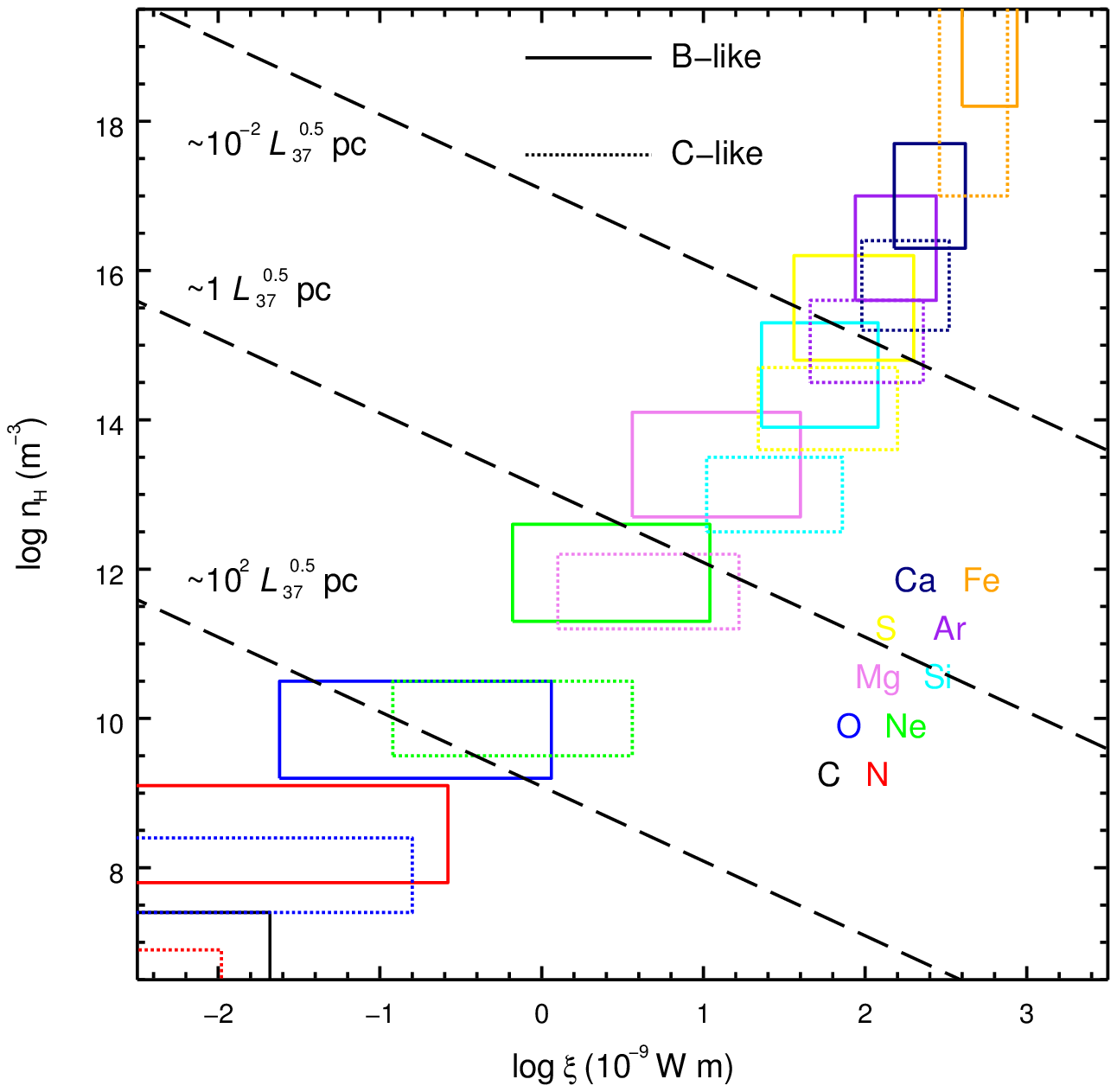}
\caption{Domain of density and ionization parameter where metastable absorption line from B-like (solid) and C-like (dotted) ions can be used for diagnostics in the case of photoionized equilibrium (see Section~\ref{sct:domain} for a  detailed description). Dashed lines indicate the distance of the photoionized plasma with respect to the central engine. $L_{37}$ is the 1 to 1000 Ryd luminosity in units of $10^{37}$~W.
}
\label{fig:ndxi_spex_dev_pion}
\end{figure}

Ions  from different isoelectronic sequences cover an extensive area in the $n_{\rm H}$ -- $\xi$ two-dimensional parameter space. In Figure~\ref{fig:ndxi_spex_dev_pion}, each box corresponds to a certain ion. Within the box, the density of the plasma can be well constrained with lines from the metastable levels. Above (below) the box, only a lower (upper) limit can be obtained. 

The box width gives the range of ionization parameters where lines from the ground and metastable levels are expected to be detected. The lower and upper boundaries correspond to $1/e$ of the maximum ion concentration using the AGN1 SED. Again we caution that a different ionizing SEDs yield different ionization parameters\footnote{Additionally, with the same ionizing SED, different photoionization codes (e.g., SPEX, XSTAR and Cloudy) yield different ionization parameters \citep[for further details, see][]{meh16}}. 

The box height corresponds to the density of the plasma. The lower and upper boundaries of the box corresponds to 10\% and 99\% of the maximum ratio of  metastable to ground level population. Only the $2s^2 2p~(^{2}P_{3/2})$ metastable level in B-like ions and the $2s^2 2p^2~(^3P_1~{\rm and}~^3P_2)$ metastable levels in C-like ions are used for the calculation since their corresponding metastable-to-ground-EW ratios barely depend on the ionization parameter (Figure~\ref{fig:ewrt_spex_dev_pion_14ins}). 

Be-like ions are not included in Figure~\ref{fig:ndxi_spex_dev_pion} because the level population depends on both the ionization parameter (thus SED dependent) and density, and because the metastable levels are less populated  than B-like and C-like ions at the
same density (Section~\ref{sct:sum}).

In addition, we also show the distance of the plasma inferred from the definition of ionization parameter (Equation~\ref{eq:xi_def}). Within the same isoelectronic sequence, high-$Z$ ions probe higher density, higher ionization parameters and smaller distances, while low-$Z$ ions measure lower density, lower ionization parameters, and larger distances. For ions in the same isonuclear sequence (i.e., the same $Z$), less ionized ions can probe lower density, thus larger distances. 

\section{Density diagnostics for the ionized outflow in NGC\,5548}
\label{sct:ngc5548}
NGC\,5548 is the archetypal Seyfert 1 galaxy. As such, it exhibits all the typical spectral features seen in type 1 Seyfert galaxies. The broadband (from optical to hard X-ray) spectral energy distribution (SED) and various properties from the multiphase ionized outflow are well studied \citep[e.g.,][]{kaa14, meh15b, ara15, ebr16}. NGC\,5548 was observed with {\it Chandra} in January 2002 with both high- and low-energy transmission grating spectrometers (HETGS and LETGS), which allow us to study the X-ray absorption features in a wide wavelength range ($\sim$2-60~\AA). Moreover, the 2002 spectra have the best signal-to-noise ratio of all the high-resolution grating spectra of NGC\,5548.

Here we reanalyze these spectra to search for density diagnostic lines. The HETGS (ObsID: 3046, with $\sim150$~ks exposure) and LETGS (ObsID: 3383 and 3045, with $\sim340$~ks total exposure) spectra are optimally binned \citep{kaa16} and fitted simultaneously. Fits to LEG, MEG, and HEG spectra are restricted to the 11--60~\AA, 4--19~\AA,  and 1.8--10~\AA\ wavelength range, respectively.

Six PION components are used to account for the six components in the ionized outflow \citep{kaa14, ebr16}. Assuming a low  plasma density ($n_{\rm H}=10^{6}~{\rm m^{-3}}$ or $1~{\rm cm^{-3}}$, kept frozen) for all six PION components, the best-fit C-statistics are 5146.0 (denoted as the baseline C-statistics in the following) with a degree of freedom of 4843. The hydrogen column density ($N_{\rm H}$), ionization parameter ($\xi$), turbulent velocity ($v_b$), and outflow velocity ($v_{\rm out}$) for the PION components (Table~\ref{tbl:5548_wax}) are consistent with values found in previous studies by \citet{kaa14} and \citet{ebr16}. 

\begin{table*}
\caption{Parameters of the six PION components in NGC\,5548.}
\label{tbl:5548_wax}
\centering
\small
\begin{tabular}{lrrrrrrrr}
\hline\hline
\noalign{\smallskip}
Component & A & B & C & D & E & F \\
\noalign{\smallskip}
\hline
\noalign{\smallskip} 
$N_{\rm H}$ ($10^{24}~{\rm m^{-2}}$) & $2.6\pm0.8$ & $6.9\pm0.9$ & $10.8\pm2.8$ & $13.4\pm2.1$ & $25\pm13$ & $52.0\pm8.5$ \\
\noalign{\smallskip} 
$\log_{10} (\xi)$ & $0.51\pm0.12$ & $1.35\pm0.06$ & $2.03\pm0.04$ & $2.22\pm0.03$ & $2.47\pm0.13$ & $2.83\pm0.03$ \\
\noalign{\smallskip} 
$v_b$ (${\rm km~s^{-1}}$) & $150\pm29$ & $49\pm14$ & $40\pm10$ & $67\pm17$ & $6\pm5$ & $115\pm29$ \\
\noalign{\smallskip} 
$v_{\rm out}$ (${\rm km~s^{-1}}$) & $-557\pm37$ & $-547\pm35$ & $-1108\pm31$ & $-271\pm24$ & $-670\pm14$ & $-1122\pm34$ \\
\noalign{\smallskip} 
$n_{\rm H}$ (${\rm m}^{-3}$) & -- -- & $\gtrsim10^{13}$ & $\lesssim10^{19}$ & $\lesssim10^{19}$ & $\lesssim10^{20}$ & $\lesssim10^{19}$ \\
\noalign{\smallskip} 
\hline
\end{tabular}
\tablefoot{The lower or upper limits of the density ($n_{\rm H}$) are at the confidence level of $\gtrsim 3\sigma$, while the statistical uncertainties of all the other parameters are at the confidence level of $1\sigma$.}
\end{table*}

\begin{figure}
\centering
\includegraphics[width=\hsize]{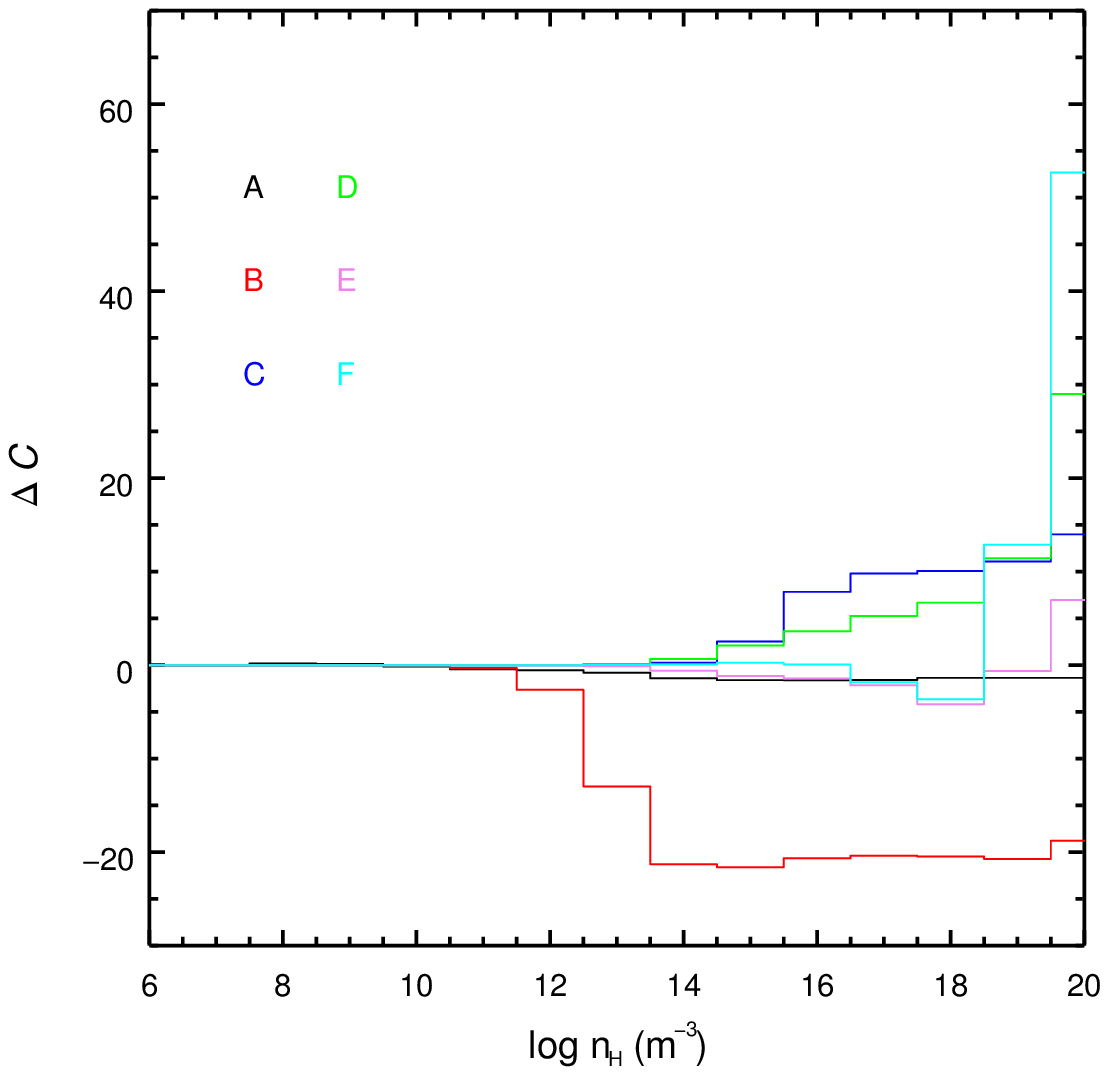}
\caption{Deviation of C-statistics ($\Delta C$) from the baseline fit (C-stat. = 5146.0 and d.o.f. = 4843) with varying plasma density ($n_{\rm H}$) for each photoionized absorber component (A--F). Component A refers to the least ionized photoionized absorber and F refers to the most ionized photoionized absorber (Table~\ref{tbl:5548_wax}).
}
\label{fig:del_cstat_hden}
\end{figure}

For each PION component, we then vary its density from $n_{\rm H}=10^{6}~{\rm m^{-3}}$ to ${10^{20}~{\rm m^{-3}}}$, with one step per decade. All the other parameters are kept frozen. The deviation of C-statistics ($\Delta C$) from the baseline fit are demonstrated in Figure~\ref{fig:del_cstat_hden}. 

For the least ionized component A, the X-ray spectra with $\lambda \lesssim 60$~\AA\ are insensitive to the density. The density sensitive lines that can be distinguished from the ground absorption lines of \ion{O}{IV} (B-like) and \ion{Ne}{V} (C-like) are at longer wavelength range ($\lambda \gtrsim 100$~\AA, Table~\ref{tbl:cal_5ies} and \ref{tbl:cal_6ies}). That is to say, high-resolution UV spectra are required to determine the density for this component.

\begin{figure}
\centering
\includegraphics[width=\hsize]{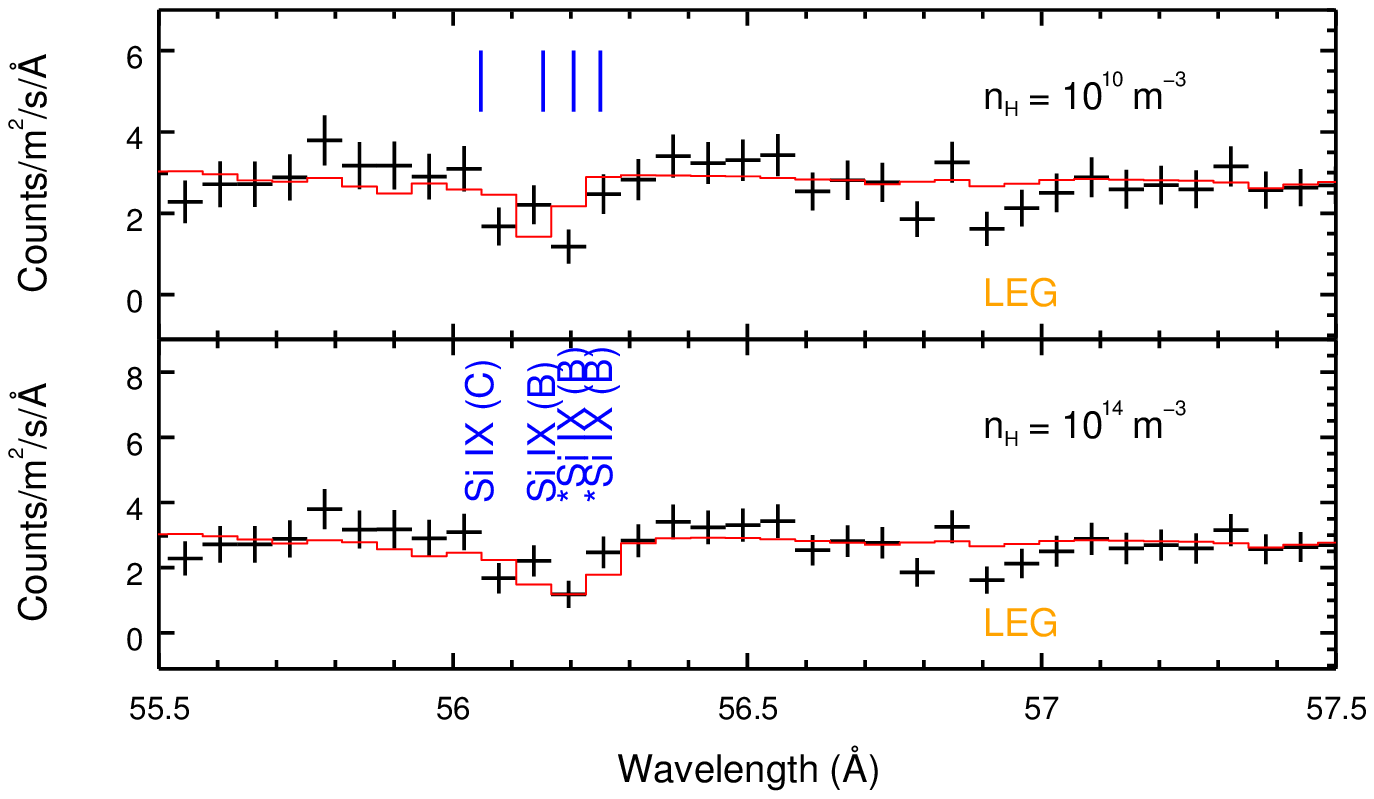}
\caption{PION modeling of the LEG spectrum of NGC\,5548 with $n_{\rm H}=10^{10}~{\rm m^{-3}}$ (upper panel) and $10^{14}~{\rm m^{-3}}$ (lower panel). The spectral bin size is 0.025~\AA. The overall C statistics is 5146.0 (d.o.f. = 4843) at low density and 5124.7 (d.o.f. = 4842) at high density. The two density sensitive metastable absorption lines of \ion{Si}{IX} (C-like), from component B, at 55.20~\AA\ and 56.25~\AA\ (in the observed frame) are labeled with $\ast$. 
}
\label{fig:spec_plot_061409_spex_dev}
\end{figure}

Meanwhile, we find a lower limit (at the confidence level of $3\sigma$) of $n_{\rm H} \gtrsim 10^{13}~{\rm m^{-3}}$ for component B. 
Figure~\ref{fig:spec_plot_061409_spex_dev} shows the LETGS spectrum in the neighborhood of \ion{Si}{IX} absorption lines ($\sim56$~\AA), where the C-statistics are improved at high density. When $n_{\rm H} \gtrsim10^{13}~{\rm m^{-3}}$, the population of the ground level $2s^2 2p^2~(^3P_{0})$ decreases, while the population of the metastable levels $2s^2 2p^2~(^3P_1~{\rm and}~^3P_2)$ increases (Figure~\ref{fig:lpop_spex_dev_pion_6ies}). Accordingly, the ground absorption line at $56.15$~\AA\ (in the observed frame) is shallower, while the metastable absorption lines at $56.20$~\AA\ and $56.25$~\AA\ are deeper. The $3\sigma$ upper limit of the distance of component B is accordingly 0.23~pc. The inferred density and distance disagree with the results reported in \citet{ebr16}, where a timing analysis (Section~\ref{sct:intro}) is used. For the photoionized absorber component B, \citet{ebr16} report $n_{\rm H}\in(2.9,~7.1)\times 10^{10}~{\rm m}^{-3}$ and $d \in (13,~20)$~pc, both at the confidence level of $1\sigma$. The upper limits of density is based on the non-detection of variability on smaller timescales. The authors also note that there is a marginal hint of recombination timescale of $\sim4$ and $\sim60$ days, which would indicate a density lower limit ($1\sigma$) of $\sim10^{10-11}~{\rm m}^{-3}$, which would agree more with our results.

\begin{figure}
\centering
\includegraphics[width=\hsize]{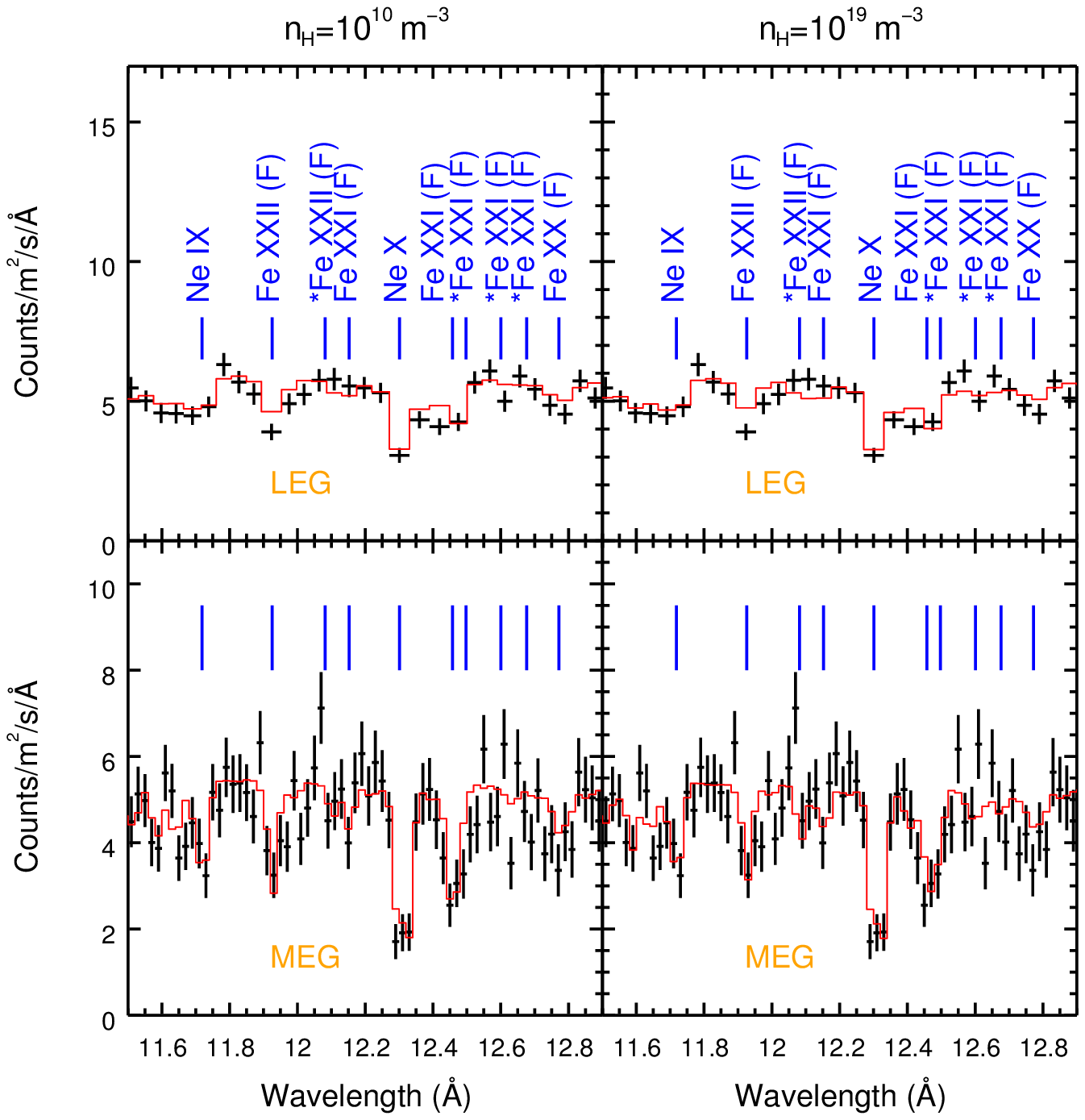}
\caption{PION modeling of the spectra of NGC\,5548 with $n_{\rm H}=10^{10}~{\rm m^{-3}}$ (left panels) and $10^{19}~{\rm m^{-3}}$ (right panels). The spectral bin sizes are 0.025~\AA\ (for LEG) and 0.01~\AA\ (for MEG), respectively. The overall C statistics is 5146.0 (d.o.f. = 4843) at low density and 5158.9 (d.o.f. = 4842) at high density. The density sensitive metastable absorption lines of \ion{Fe}{XXII} (B-like) and \ion{Fe}{XXI} (C-like), from component F, are labeled with $\ast$.
}
\label{fig:spec_plot_05to0626_spex_dev}
\end{figure}

For components C -- F, upper limits are obtained with $n_{\rm H} \lesssim10^{19}~{\rm m^{-3}}$ (above $3\sigma$ for components C, E, and F) and $\lesssim10^{20}~{\rm m^{-3}}$ ($2.6\sigma$ for component D). We show in Figure~\ref{fig:spec_plot_05to0626_spex_dev} the spectra in the neighborhood of \ion{Fe}{XXII} (B-like) and \ion{Fe}{XXI} (C-like) with the density of component F set to $n_{\rm H}=10^{10}~{\rm m^{-3}}$ (left panels) and $10^{19}~{\rm m^{-3}}$ (right panels). At high density the metastable absorption lines at 12.08~\AA\ (\ion{Fe}{XXII}), at 12.50~\AA\ (\ion{Fe}{XXI}), and at 12.60~\AA\ (\ion{Fe}{XXI}) are deeper. This overestimation of absorption lines contradicts  the data and leads to  poorer C-statistics. Of course, the spectra are crowded in this wavelength range, so that density diagnostics are challenging. The $3\sigma$ lower limits of the distance of components C -- F are a few light-hours ($\sim10^{12}~{\rm m}$). The obtained density and distance for components C -- F do not contradict the results ($n_{\rm H} \gtrsim10^{10}~{\rm m^{-3}}$ and $d\lesssim 1$~pc) reported in \citet{ebr16}.   

Due to the narrow wavelength coverage and rather limited effective area of current grating instruments, and the lack of atomic data for N-like to F-like ions, density diagnostics using absorption lines from the metastable levels are not very effective. Once the N-like to F-like atomic data are included, a significant portion of the $n_{\rm H}$ -- $\xi$ parameter space can be covered. This, combined with the next generation of spectrometers on board {\it Arcus} \citep{smi16} and Athena \citep{nan13},  will allow us to identify the presence/absence of these density-sensitive absorption lines \citep{kaa17b}, thus tightly constraining the location and the kinetic power of AGN outflows. We refer the readers to Fig.~6 of \citet{kaa17b} for a simulated {\it Arcus} spectrum of NGC\,5548, compared with the observed 2002 LETGS spectrum. 

\begin{acknowledgements}
We thank the referee for the constructive comments and suggestions. SRON is supported financially by NWO, the Netherlands Organization for Scientific Research.
\end{acknowledgements}


\begin{appendix}
\section{N-like \ion{Fe}{XX} to F-like \ion{Fe}{XVIII}}
\label{sct:07to09ies}
\begin{figure}
\centering
\includegraphics[width=\hsize]{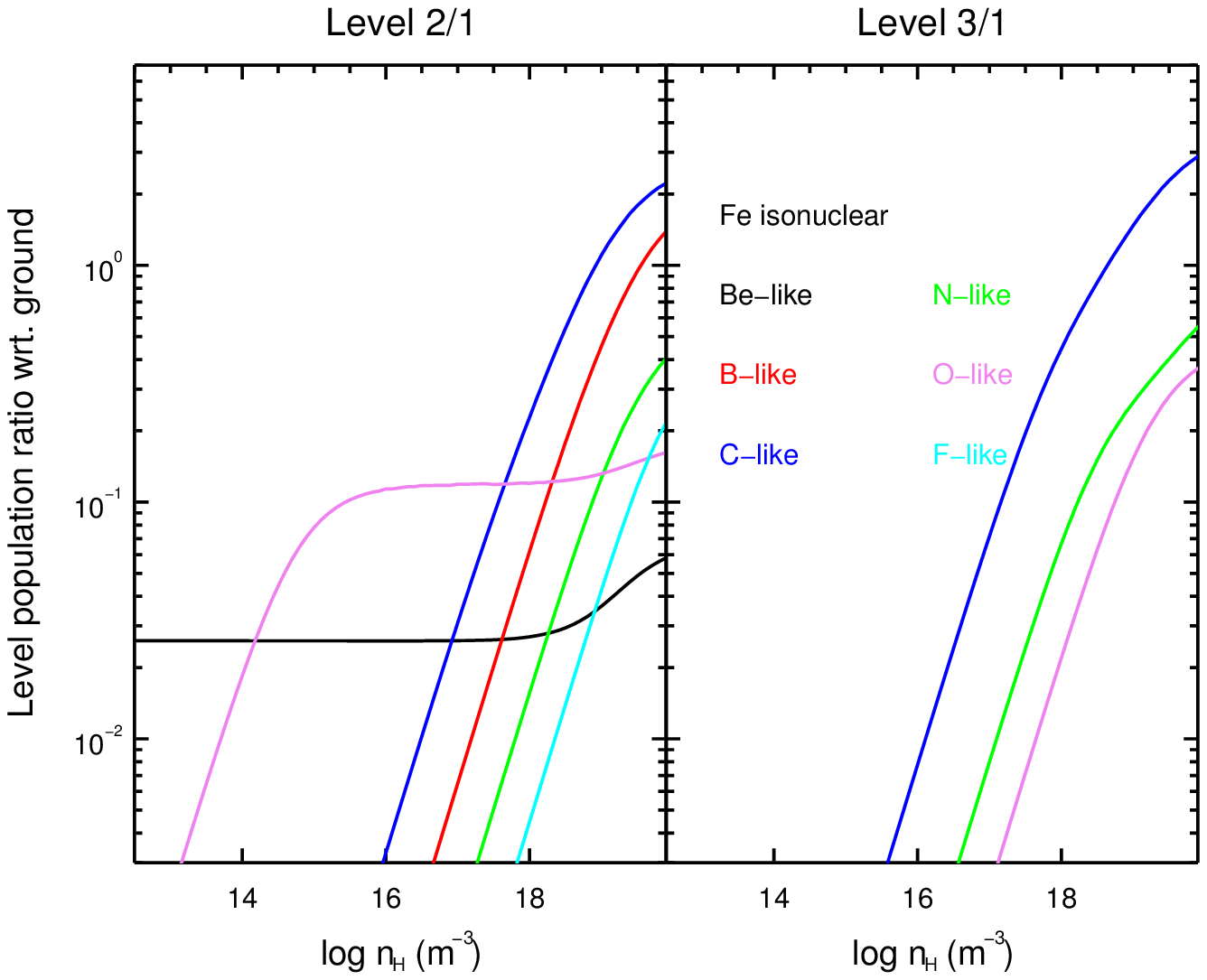}
\caption{Ratio of metastable to ground level population  as a function of density of the Fe isonuclear sequence (from Be-like to F-like) at the temperature of maximum ion concentration in ionization equilibria. The configuration and $^{2S +1}L_{J}$ notation of the ground (Level 1) and metastable levels (Level 2 and 3) are listed in Table~\ref{tbl:lev_idx}.}
\label{fig:lpop_spex_dev_pion_26ins}
\end{figure}

Currently, in SPEXACT v3.04, the atomic data for the N-like to F-like isoelectronic sequences are lacking, except for \ion{Fe}{XX}, \ion{Fe}{XIX}, and \ion{Fe}{XVIII}. Here we repeat the same analyses described above, but for N-like to F-like Fe. For both N-like \ion{Fe}{XX} and O-like \ion{Fe}{XIX}, the first two excited levels can be populated up to 50\%, compared to the ground level population, while only the first excited level of F-like \ion{Fe}{XVIII} can be populated significantly (Figure~\ref{fig:lpop_spex_dev_pion_26ins}). Characteristic transitions from these ground and metastable levels are listed in Table~\ref{tbl:cal_26ins}. 
\begin{table*}
\caption{Characteristic absorption lines from the ground and metastable levels in N-like to F-like iron ions.}
\label{tbl:tran_26ins}
\centering
\small
\begin{tabular}{ccllllllllll}
\hline\hline
\noalign{\smallskip}
\multicolumn{2}{r}{Index} & \multicolumn{2}{l}{1} &  \multicolumn{2}{l}{2} &  \multicolumn{2}{l}{3}   \\
\noalign{\smallskip}
\hline\hline
\noalign{\smallskip}
Sequence & $n_j$ & Lower & Upper & Lower & Upper & Lower & Upper \\
\noalign{\smallskip}
\hline
\noalign{\smallskip}
\ion{Fe}{XX} & 2-2 & $2s^2 2p^3~(^4S_{3/2})$ & $2s 2p^4~(^4P_{5/2})$ & $2s^2 2p^3~(^2D_{3/2})$ & $2s 2p^4~(^2D_{3/2})$ & $2s^2 2p^3~(^2D_{5/2})$ & $2s 2p^4~(^2P_{3/2})$  \\
\ion{Fe}{XX} & 2-3 & $2s^2 2p^3~(^4S_{3/2})$ & $2s^2 2p^2 (^3P) 3d~(^4P_{3/2})$ & $2s^2 2p^3~(^2D_{3/2})$ & $2s^2 2p^2 (^3P) 3d~(^2D_{5/2})$ & $2s^2 2p^3~(^2D_{5/2})$ & $2s^2 2p^2 (^3P) 3d~(^2F_{3.5})$   \\
\ion{Fe}{XX} & 1-2 & $2s^2 2p^3~(^4S_{3/2})$ & $1s 2s^2 2p^4~(^4P_{5/2})$ & $2s^2 2p^3~(^2D_{3/2})$ & $1s 2s^2 2p^4~(^2D_{3/2})$ & $2s^2 2p^3~(^2D_{5/2})$ & $1s 2s^2 2p^4~(^2D_{5/2})$  \\
\noalign{\smallskip}
\ion{Fe}{XIX} & 2-2 & $2s^2 2p^4~(^3P_{2})$ & $2s 2p^5~(^3P_{2})$ & $2s^2 2p^4~(^3P_{0})$ & $2s 2p^5~(^3P_{1})$ & $2s^2 2p^4~(^3P_{1})$ & $2s 2p^5~(^3P_{2})$  \\
\ion{Fe}{XIX} & 2-3 & $2s^2 2p^4~(^3P_{2})$ & $2s^2 2p^3 (^2D) 3d~(^3D_{3})$ & $2s^2 2p^4~(^3P_{0})$ & $2s^2 2p^3 (^2P) 3d~(^3P_{1})$ & $2s^2 2p^4~(^3P_{1})$ & $2s^2 2p^3 (^2D) 3d~(^3D_{2})$   \\
\ion{Fe}{XIX} & 1-2 & $2s^2 2p^4~(^3P_{2})$ & $1s 2s^2 2p^5~(^3P_{2})$ & $2s^2 2p^4~(^3P_{0})$ & $1s 2s^2 2p^5~(^3P_{1})$ & $2s^2 2p^4~(^3P_{1})$ & $1s 2s^2 2p^5~(^3P_{2})$  \\
\noalign{\smallskip}
\ion{Fe}{XVIII} & 2-2 & $2s^2 2p^5~(^2P_{3/2})$ & $2s 2p^6~(^2S_{1/2})$ & $2s^2 2p^5~(^2P_{1/2})$ & $2s 2p^6~(^2S_{1/2})$ & -- -- & -- -- \\
\ion{Fe}{XVIII} & 2-3 & $2s^2 2p^5~(^2P_{3/2})$ & $2s^2 2p^4 (^1D) 3d~(^2D_{5/2})$ & $2s^2 2p^5~(^2P_{1/2})$ & $2s^2 2p^4 (^1S) 3d~(^2D_{3/2})$ & -- -- & -- --\\
\ion{Fe}{XVIII} & 1-2 & $2s^2 2p^5~(^2P_{3/2})$ & $1s 2s^2 2p^6~(^2S_{1/2})$ & $2s^2 2p^5~(^2P_{1/2})$ & $1s 2s^2 2p^6~(^2S_{1/2})$ & -- -- & -- --  \\
\hline
\end{tabular}
\end{table*}
\begin{table*}
\caption{Characteristic absorption lines from the ground and the metastable levels in N-like \ion{Fe}{xx} to F-like \ion{Fe}{xviii} (SPEXACT v3.04).}
\label{tbl:cal_26ins}
\centering
\small
\begin{tabular}{llllllllll}
\hline\hline
\noalign{\smallskip}
\multicolumn{2}{l}{Index} & \multicolumn{2}{l}{1} &  \multicolumn{2}{l}{2} &  \multicolumn{2}{l}{3}  \\
\noalign{\smallskip}
\hline\hline
\noalign{\smallskip}
Ion & $n_j$ & $\lambda$~(\AA) & $f$ & $\lambda$~(\AA) & $f$ & $\lambda$~(\AA) & $f$ \\
\noalign{\smallskip}
\hline
\noalign{\smallskip} 
\ion{Fe}{XX} & 2-2 & 132.850 & 0.05 & 110.626 & 0.08 & 93.782 & 0.09 \\ 
\ion{Fe}{XX} & 2-3 & 12.835 & 0.48 & 12.984 & 0.60 & 12.885 & 0.87 \\ 
\ion{Fe}{XX} & 1-2 &1.908 & 0.15 & 1.906 & 0.24 & 1.906 & 0.15 \\ 
\noalign{\smallskip}
\ion{Fe}{XIX} & 2-2 & 108.355 & 0.06 & 109.952 & 0.08 & 119.983 & 0.03 \\ 
\ion{Fe}{XIX} & 2-3 &13.521 & 0.73 & 13.420 & 1.23 & 13.735 & 0.50 \\ 
\ion{Fe}{XIX} & 1-2 &1.918 & 0.15 & 1.918 & 0.23 & 1.921 & 0.10 \\ 
\noalign{\smallskip}
\ion{Fe}{XVIII} & 2-2 &93.923 & 0.06 & 103.937 & 0.05 & -- -- & -- -- \\ 
\ion{Fe}{XVIII} & 2-3 &14.203 & 0.88 & 14.121 & 0.86 & -- -- & -- -- \\ 
\ion{Fe}{XVIII} & 2-1 &1.928 & 0.11 & 1.931 & 0.11 & -- -- & -- -- \\ 
\hline
\end{tabular}
\end{table*}

\begin{figure}
\centering
\includegraphics[width=\hsize]{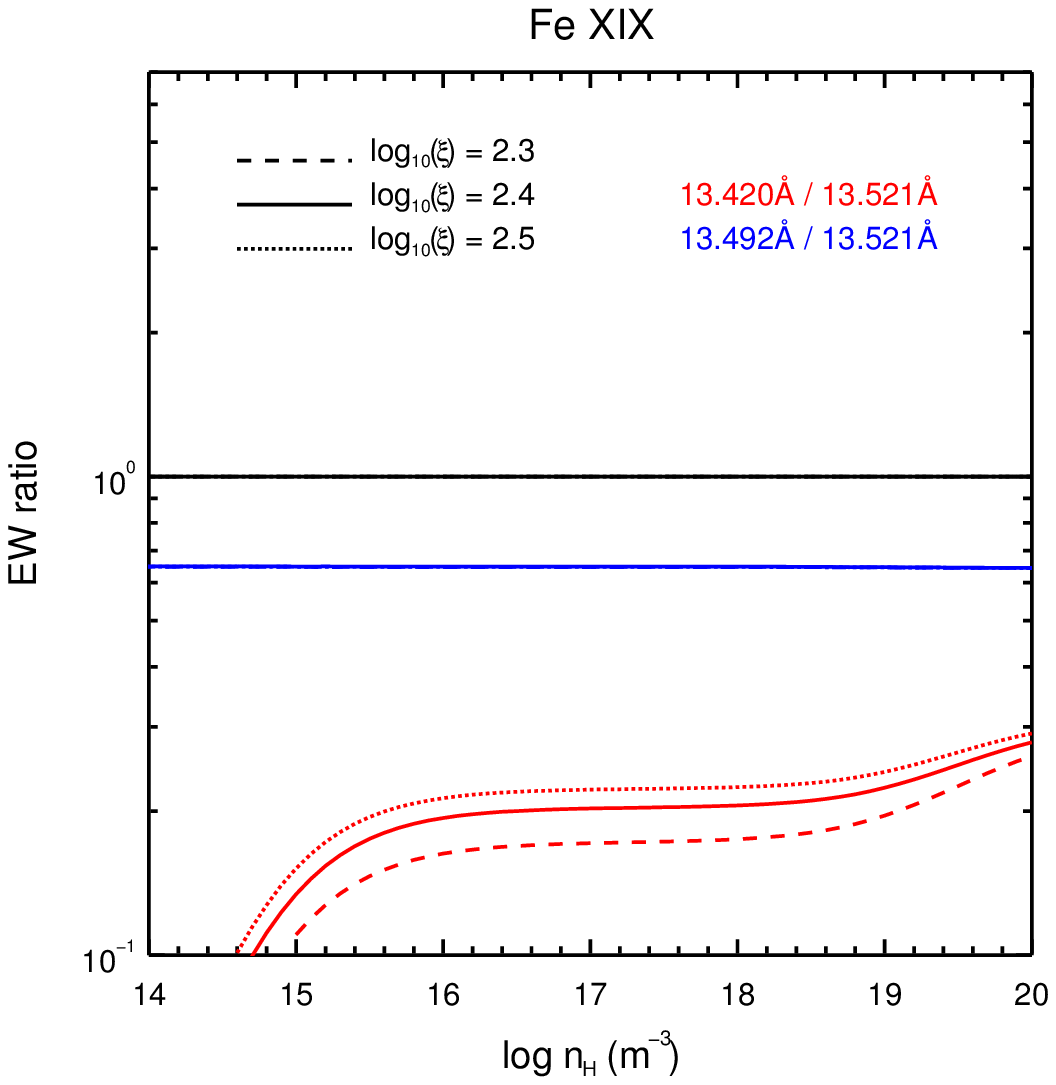}
\caption{Equivalent width (EW) ratios for characteristic absorption lines of O-like \ion{Fe}{XIX}. The two lines with $\lambda=$13.521~\AA\ and 13.492~\AA\ are
from the ground level. The line with $\lambda=$13.420~\AA\ is from the first excited level. Dashed ($\xi = 2.3$), solid ($\xi = 2.4$, where ion concentration reaches its maximum), and dotted ($\xi = 2.5$) lines indicate different ionization parameters of the photoionized plasma.}
\label{fig:ewrt_spex_dev_pion_8_26}
\end{figure}

Comparing the density-sensitive metastable levels for Be-like \ion{Fe}{XXIII} to F-like \ion{Fe}{XVIII} (Figure~\ref{fig:lpop_spex_dev_pion_26ins}), we find that, in general, the metastable levels in C-like \ion{Fe}{XXI} can be more easily populated when $n_{\rm H}\gtrsim10^{16}~{\rm m^{-3}}$. In addition, when $n_{\rm H}\lesssim10^{16}~{\rm m^{-3}}$, the first excited level $2s^2~2p^4~(^3P_0)$ of O-like \ion{Fe}{XIX} can be populated up to 10\% of the ground level population, which has been pointed out in \citet{kas01}. Nonetheless, the density-sensitive line (13.420~\AA) of O-like \ion{Fe}{XIX} is weaker than the ground level absorption lines (13.521~\AA\ and 13.492~\AA). Meanwhile, the density-sensitive line (13.420~\AA) of O-like \ion{Fe}{XIX} is sensitive to both ionization parameter and density (Figure~\ref{fig:ewrt_spex_dev_pion_8_26}). Additionally, from the observational point of view, we caution that the $n_j=2-3$ transitions of \ion{Fe}{XVIII} at 14.203~\AA\ (ground) and 14.121~\AA\ (metastable) might be blended with the \ion{O}{VIII} absorption edge at 14.228~\AA.  

\section{Comparison of the level population calculation with CHIANTI}
\label{sct:cf_chianti}
CHIANTI \citep{dza15} is a widely used atomic code for analyzing  emission line spectra from astrophysical sources. Similar to SPEX, CHIANTI also provides  detailed calculations (e.g., level populations). Therefore, here we compare the level population calculations for collisional ionized equilibrium  (CIE) plasmas, using the latest version of CHIANTI v8.0.8 (with atomic database v8.0.6\footnote{http://www.chiantidatabase.org/}) and SPEX v3.04 (with SPEXACT v3.04) to show the effects of different atomic codes.

\begin{figure*}
\centering
\includegraphics[width=\hsize, trim={1cm 2cm 1cm 2cm}, clip]{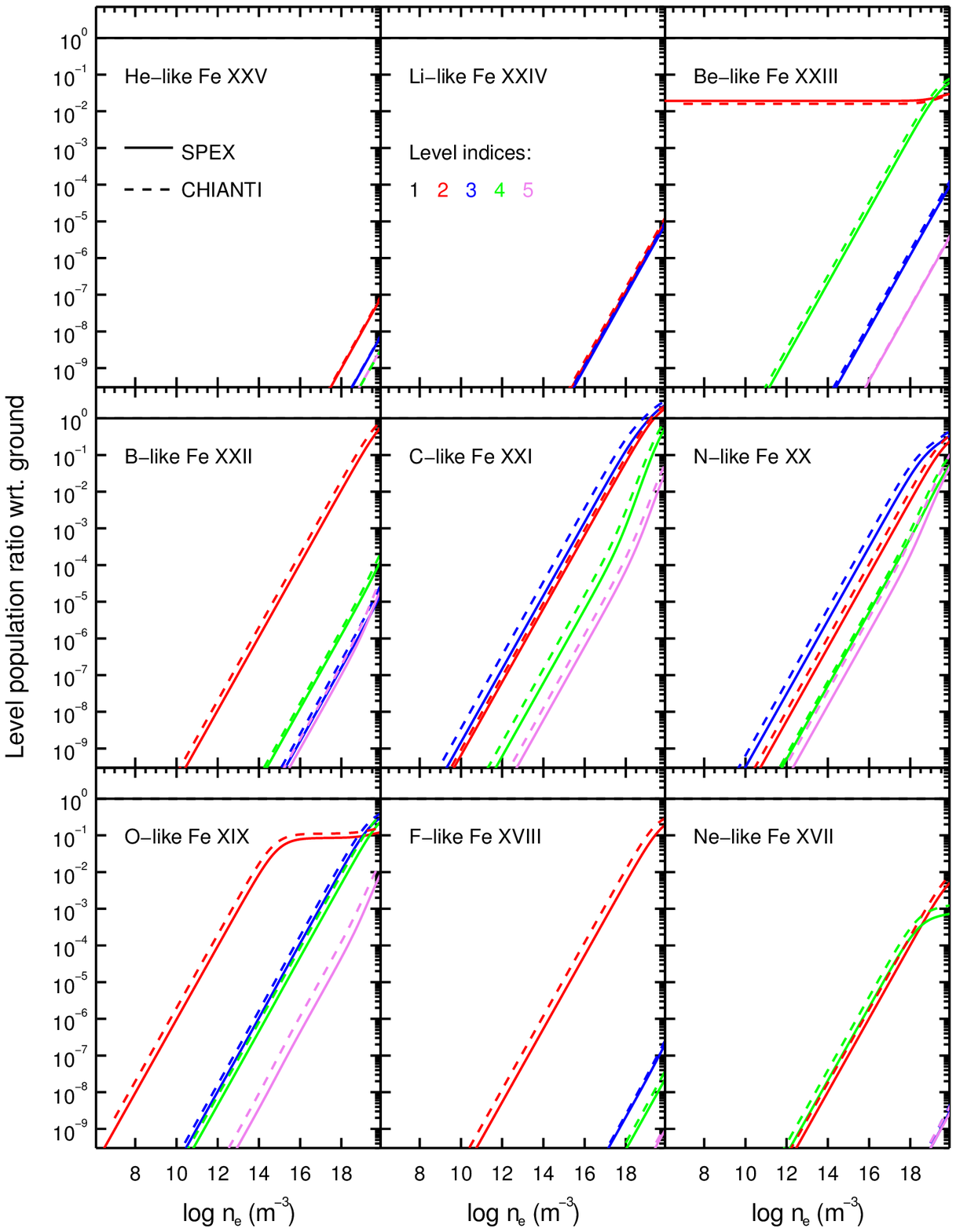}
\caption{Level population ratio for the first four excited levels (level indices 2--5) of the Fe isonuclear sequence from He-like \ion{Fe}{XXV} to Ne-like \ion{Fe}{XVII} for collisional ionized equilibrium (CIE) plasma, using SPEXACT v3.04 (solid line) and CHIANTI v8.0.8 (dashed line). The CHIANTI atomic database v8.0.6 is used. }
\label{fig:pop_cie_cf_26ins}
\end{figure*}

For simplicity, we only compare the population of the first five levels (i.e., the ground level and the first four excited levels) of the Fe isonuclear sequence from He-like \ion{Fe}{XXV} to Ne-like \ion{Fe}{XVII}. The level populations are calculated in both codes using the same abundance table \citep{lod09}, the same ionization balance \citep{bry09}, and the same plasma temperature (where the ion concentration reaches its maximum). Figure~\ref{fig:pop_cie_cf_26ins} shows the level population ratio with respect to the ground as a function of plasma density ranging from $10^{6}~{\rm m^{-3}}$ to $10^{20}~{\rm m^{-3}}$ (or $10^{1}~{\rm cm^{-3}}$ to $10^{14}~{\rm cm^{-3}}$).

For He-like \ion{Fe}{XXV} and Li-like \ion{Fe}{XXIV}, the level population ratios provided by the two codes are consistent with each other. For Be-like \ion{Fe }{XXIII} to Ne-like \ion{Fe}{XVII}, the level population ratios share the same increasing and/or flattening trend, whereas the two codes yield slightly different level population ratios (a factor of few). This is not totally unexpected, given the fact that the two codes do not use the same atomic data for individual atomic processes, and the two codes use different means of implementing the atomic data for the calculation (interpolation or parameterization). For instance, the CHIANTI atomic database includes various atomic data with $n\le5$ for almost all ions with $Z\le30$ in the He-like to Ne-like isoelectronic sequences, while the SPEXACT includes atomic data with $n\le8$ for He-like to C-like ions with $Z\le30$, and only Fe ions in the N-like to Ne-like isoelectronic sequences. The atomic data collected in both codes are state-of-the-art, yet still incomplete and not 100\% accurate. Furthermore, the SPEX code parameterizes almost all the atomic data \citep[e.g.,][]{mao16,urd17}, while CHIANTI uses interpolation instead.

In short, considering only the systematic uncertainties introduced by the different atomic data used in SPEX and CHIANTI, with the same density diagnostics in Be-like to Ne-like ions, the two codes are expected to estimate the plasma density to a similar order of magnitude. However, it is not clear which part(s) of the atomic data and/or code contribute to what percentage of the total ``error'' budget  (i.e., the discrepancy between the codes)  for the level population calculation of a specific level in a specific ion.

\end{appendix}

\end{document}